\documentclass[twocolumn]{aastex63}

\usepackage{scalerel}

\newcommand\orcidicon[1]{\href{https://orcid.org/#1}{\mbox{\scalerel*{\includegraphics{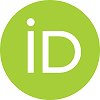}}{|}}}}

\usepackage[]{hyperref}
\usepackage{amsmath}
\usepackage[]{graphicx}
\newcommand{\theTitle}{The Effect of Solar Wind Expansion and Non-Equilibrium Ionization \\ on
	the Broadening of Coronal Emission Lines}
\newcommand{\theShortTitle}{Line of Sight Spectroscopy in the Solar Corona}

\hypersetup{
	pdftitle={\theTitle},
	pdfauthor={Chris R. Gilly},
	pdfsubject={\theShortTitle},
	pdfkeywords={Sun: oscillations -- Sun: atmosphere -- Sun: corona -- solar wind -- Sun: UV radiation},
	bookmarksnumbered=true,     
	bookmarksopen=true,         
	bookmarksopenlevel=1,       
	colorlinks=true,            
	pdfstartview=Fit,           
	pdfpagemode=UseOutlines,   
	breaklinks=true,
	urlcolor=blue,
	citecolor=blue,
	linkcolor=blue}

\shorttitle{\theShortTitle}
\shortauthors{Gilly \& Cranmer}

\newcommand{\linewidthb}{1.02\linewidth} %All the plots are this wide
\newcommand{\enn}{{\bf \hat{n}}}
\newcommand{\appropto}{\mathrel{\vcenter{\offinterlineskip\halign{\hfil$##$\cr\propto\cr\noalign{\kern2pt}\sim\cr\noalign{\kern-2pt}}}}}

\received{06/17/2020}
\revised{07/28/2020}
\accepted{08/20/2020}
\begin{document}
	
\title{\theTitle}

\author[0000-0003-0021-9056]{Chris R. Gilly \orcidicon{0000-0003-0021-9056}}
%\affiliation{Laboratory for Atmospheric and Space Physics}/
\affiliation{LASP; Department of Astrophysical and Planetary Sciences, University of Colorado, Boulder, Colorado, 80309, USA}

\author[0000-0002-3699-3134]{Steven R. Cranmer \orcidicon{0000-0002-3699-3134}}
%\affiliation{Laboratory for Atmospheric and Space Physics}
\affiliation{LASP; Department of Astrophysical and Planetary Sciences, University of Colorado, Boulder, Colorado, 80309, USA}

%\submitjournal{\apj} \received{6/17/2020}

\begin{abstract}
	
When observing spectral lines in the optically-thin corona, line-of-sight (LOS) effects can strongly affect the interpretation of the data, especially in regions just above the limb. We present a semi-empirical forward model, called GHOSTS, to characterize these effects. GHOSTS uses inputs from several other models to compute non-equilibrium ionization states (which include the solar-wind freezing-in effect) for many ions. These are used to generate ensembles of simulated spectral lines that are examined in detail, with emphasis on: (1) relationships between quantities derived from observables and the radial variation of the observed quantities, (2) the behavior of thermal and non-thermal components of the line width, and (3) relative contributions of collisionally excited and radiatively scattered photons.
We find that rapidly changing temperatures in the low corona can cause ion populations to vary dramatically with height. This can lead to line-width measurements that are constant with height (a ``plateau'' effect) even when the temperature is increasing rapidly, as the plane-of-sky becomes evacuated and the foreground/background plasma dominates the observation.
We find that LOS effects often drive the velocity width to be close to the plane-of-sky value of the wind speed, despite it flowing perpendicularly to the LOS there. The plateau effect can also cause the non-thermal component of the line width to greatly exceed the solar wind velocity at the observation height. 
Lastly, we study how much of the LOS is significant to the observation, and the importance of including continuum in the solar spectrum when computing the radiatively scattered emission.

\end{abstract}


%\section*{}
%\newpage
% \pagebreak
%\tableofcontents

%%%%%%%%%%%%%%%%%%%%%%%%%%%%%%%%%%%%%%%%%%%%%%%%%%%%%%%%%%%%%%%%%%%%%%%%%%%%%%%%%%%%%%%%%%%%%%%%%%%%%%%%%%%%%%%%%%%%%%%%%%%%%%%%%%%%%%%%%%%%%%%%%%
%%%%%%%%%%%%%%%%%%%%%%%%%%%%%%%%%%%%%%%%%%%%%%%%%%%%%%%%%%%%%%%%%%%%%%%%%%%%%%%%%%%%%%%%%%%%%%%%%%%%%%%%%%%%%%%%%%%%%%%%%%%%%%%%%%%%%%%%%%%%%%%%%%

\section{Introduction and Motivation} \label{sec:intro}

	Spectroscopy is a powerful tool for determining conditions in the solar corona. With it, we can learn about the temperatures, densities, velocity distributions, and abundances of electrons, protons, and minor ions near the Sun \citep[see, e.g.,][]{Withbroe1982, Kohl2006, Slemzin2014, DelZanna2018}. But the interpretation of these lines requires care. Above the solar limb, the corona becomes transparent, or optically thin, and observed spectral lines consist of light from a range of points with very different conditions, along an extended line of sight (LOS). Because the density of the corona drops off rapidly with altitude, and since the plane-of-the-sky (POS) is the closest point to the sun along a given LOS, the POS is assumed to be the most dense (and therefore brightest) structure sampled at a given observation height. It is straightforward, then, to assume that a measurement taken at an observation height $b$ of 2 solar radii ($R_\odot$) off the limb should be dominated by plasma at or around 2 $R_\odot$ above the solar surface. However, when making spectral measurements, the POS can not always be considered to be dominant, as the density of a given emitting ion changes rapidly with temperature. We believe that a forward model is necessary to explore some of the outstanding questions about the impact of these so-called LOS effects on the observations. 
	
	Many aspects of these observations remain poorly understood in the literature. As a representative example, there have been several measurements taken of anomalously broad lines of O VI around $b=2R_\odot$ \citep{Kohl1997, Cranmer1999, Esser1999, Kohl1999}. This has been interpreted to indicate ion temperatures in excess of $10^8$ Kelvin, which is significantly higher than the local electron temperature. Yet recent observations of spectral widths in the lower corona, with $b$ of $1.2$ to $1.5 R_\odot$ above the solar center, have been narrower than expected \citep{Hahn2013}, implying that outward flowing Alfv\'en waves may be damping out faster than predicted. This makes it harder to explain the source of the energy for the differential acceleration and preferential heating of ions. These processes are expected to occur, but the extent and precise mechanisms involved are unclear.
	
	This is a golden era for solar observations, and many new observatories are activating all over the world and in space, such as the \textit{Daniel K Inoyue Solar Telescope} (DKIST), \textit{Solar Orbiter} (SO), the \textit{Parker Solar Probe} (PSP), as well as scores of smaller missions. In order to understand what these instruments are telling us, it is important that a detailed study of line-of-sight effects be carried out, to build on the considerable work already undertaken in the literature \citep[see, e.g.,][]{Judge2007, Kohl2008, Gibson2016, VanDoorsselaere2016, Vourlidas2018, Laming2019, Zhao2019}.

	In this work, we present the Global Heliospheric Optically-thin Spectral Transport Simulation (GHOSTS). GHOSTS is a semi-empirical model, which uses inputs from several other models to generate ensembles of simulated observations for arbitrary lines of sight through the model corona assuming optically-thin radiative transfer. Non-equilibrium ionization calculations are performed on the input parameters to help determine the observables. GHOSTS is able to operate as a slit spectrograph or as a spectral imager, with the ability to evolve the observation in time. An advantage that GHOSTS has over traditional simulation methods is that the we have direct control over the properties of the physics, such as the solar wind and magnetic field strength and direction, the density and temperature as a function of space, as well as the presence of Alfv\'en waves, which can be turned on and off to examine their effects on the spectral lines. While we would like to match the real conditions of the Sun as closely as possible, the primary goal of this work is to improve the interpretation of spectral measurements, which involves comparing the inputs and outputs of the model. Spectral lines are examined to explore things such as the relative contributions of collisionally excited photons and radiatively scattered photons, the behavior of the thermal and non-thermal components of the spectral line width, and the correlation between simulated derived observables and the true radial variation of the target quantity.
	
	We begin by modeling a time-steady and axisymmetric polar coronal hole. Section \ref{sec:backgroundPlasma} describes the plasma physics that we use for the coronal-hole. Section \ref{sec:radiative} then describes the radiative transfer and spectral analysis procedures. Section \ref{sec:allResults} presents the results of the analysis, exploring the effect of the solar wind and of preferential ion heating. Finally, Section \ref{sec:discussion} provides some discussion and analysis of this work, as well as some recommendations to observers. We leave a treatment of Alfv\'en waves and other time-dependent non-thermal line-broadening to a future paper.

%%%%%%%%%%%%%%%%%%%%%%%%%%%%%%%%%%%%%%%%%%%%%%%%%%%%%%%%%%%%%%%%%%%%%%%%%%%%%%%%%%%%%%%%%%%%%%%%%%%%%%%%%%%%%%%%%%%%%%%%%%%%%%%%%%%%%%%%%%%%%%%%%%
\section{Time-Steady Plasma Physics} \label{sec:backgroundPlasma}

	Here we present the details of the physical models we used as inputs to GHOSTS, as well as the time-steady physics that we used to construct a polar coronal hole like those seen at solar minimum. Section \ref{sec:plasmaParameters} discusses plasma parameters, which were provided by the ZEPHYR model, Section \ref{sec:eqChargeStates} describes how we modeled non-equilibrium ionization states for each of the elements, and Section \ref{sec:superradialExpansion} discusses the details of the polar magnetic geometry.

\subsection{Plasma Parameters: The ZEPHYR Model} \label{sec:plasmaParameters}

    The ZEPHYR code produces a self-consistent model of the photosphere, chromosphere, corona, and solar wind \citep{Cranmer2007}. This code considers a one-dimensional open magnetic flux tube, rooted in the solar photosphere, and it calculates time-independent solutions to the hydrodynamic conservation equations with a steady turbulent heating. A notable simplification of this model is that it is single-fluid, treating protons, electrons, and ions as if they had the same velocities and temperatures.
    
\begin{figure}[ht!]
    \begin{center}
    \includegraphics[width=\linewidthb]{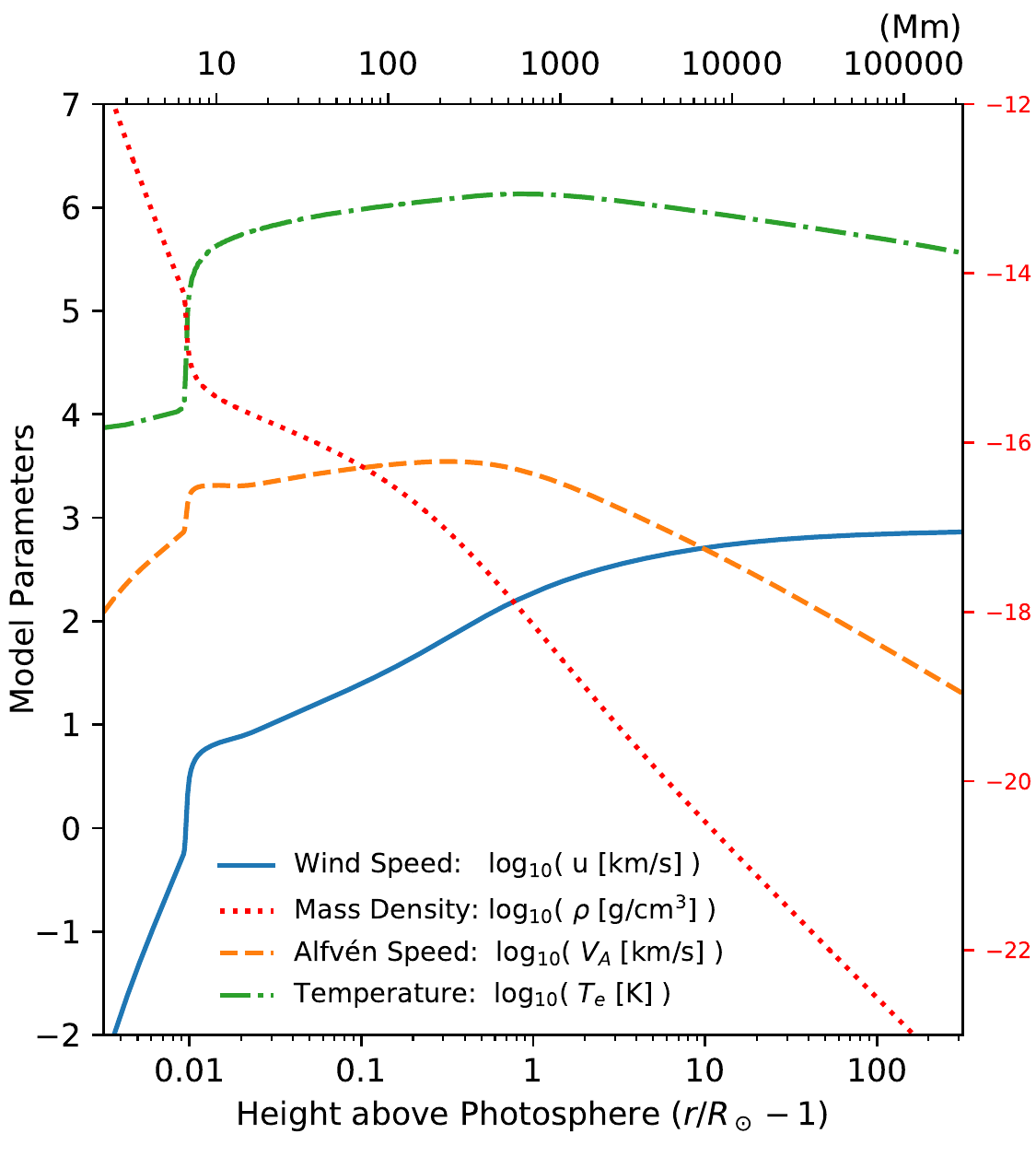}
    \caption{Tabulated output from ZEPHYR showing steady state background plasma parameters. All lines use the left scale bar except for density, which uses the right scale. }
    \label{fig:zephyr}
    \end{center}
    \end{figure}

    For this work, tabulated output was used for the mass density $\rho$, radial magnetic field strength $B$, solar wind speed $u$, and the electron temperature $T_e$, as a function of radius $r$ for a magnetic flux tube rooted at the center of a polar coronal hole (see Figure \ref{fig:zephyr}). The Alfv\'en speed was calculated using 
    \begin{equation}
    V_A = \frac{B}{\sqrt{4\pi\rho}}.
    \end{equation}
    These plasma properties are treated as steady-state background values for the entire coronal hole, which can then be perturbed locally.

\subsection{NEI Charge State Calculation} \label{sec:eqChargeStates}

	The ZEPHYR model provides the total coronal mass density as a function of height, but it does not detail the elemental abundances or nonequilibrium ionization (NEI) charge states. We used coronal elemental abundances $A_Z = n_Z/n_H$ from \citet{Schmelz2012a}, where $n_Z$ is the total number density of an element with atomic number $Z$, and $n_H$ is the total number density of hydrogen. 
	
	The NEI charge states must be computed as a function of the plasma conditions in the corona. We first found equilibrium values, which serve as an initial condition for and comparison to the more precise treatment described below. Equilibrium charge states were computed by balancing each ion's temperature-dependent collisional ionization and recombination rates, acquired from CHIANTI version 8 \citep{Dere1997, DelZanna2015}. The charge-state fractions are defined by
		
    \begin{equation}
        \frac{n_i}{n_{i-1}} = \frac{C_{i-1}}{R_{i}},
    \end{equation}
    
    \noindent where $C_i(T_e)$ is the rate at which particles of state $i$ are ionized into state $i+1$, and $R_i(T_e)$ is the rate at which particles of state $i$ recombine with an electron and fall down to state $i-1$. With the additional constraint of
    \begin{equation}
    	\sum_i {n_i} = n_{Z} = A_Z \frac{\rho}{m_p}
    \end{equation}
      
    \noindent (which just states that the sum of the ionization states must equal the total population), these equations allow for the solution of all charge state populations for an arbitrary element as a function of temperature \citep[see, e.g.,][]{Arnaud1985, Mazzotta1998}. 
    
	The charge states in the corona are only in equilibrium when the ions have time to collisionally couple with the local electron distribution before they are swept away by the solar wind \citep{Owocki1983,Esser2002,Landi2012, Boe2018}. Because the solar wind velocity increases with height and the density drops rapidly, a ``freeze-in'' radius can be defined, above which the charge states no longer have time to evolve with the local electron temperature. We follow \cite{Landi2012c} and define $R_{fr}$ as the heliocentric radial distance at which the ion fraction comes within 10\% of the asymptotic 
	frozen-in value at the maximum modeled height of $r=50R_\odot$. To model this behavior we solve the following time-steady mass conservation equations for each species,

    \begin{equation} \label{eq:rawIonization}
        \frac{1}{fr^2}\frac{\partial}{\partial r}(fr^2 n_i u) = n_e I_i ,
    \end{equation}
    \noindent where 
    \begin{equation}\label{eq:ionTerms}
		I_i = n_{i-1}C_{i-1} + n_{i+1}R_{i+1} - n_{i}(C_{i} + R_{i}),
    \end{equation}
     $n_e = \rho/m_p$ is the electron density (which assumes that hydrogen is fully ionized and neglects the 5-10\% correction due to helium), $f$ is the superradial expansion factor as described in the next section, and $i$ runs from $i=1$ (neutral) through $i=Z+1$ (fully ionized).
    
\begin{figure}[ht!]
	%    	\begin{center}
	\includegraphics[width=\linewidthb]{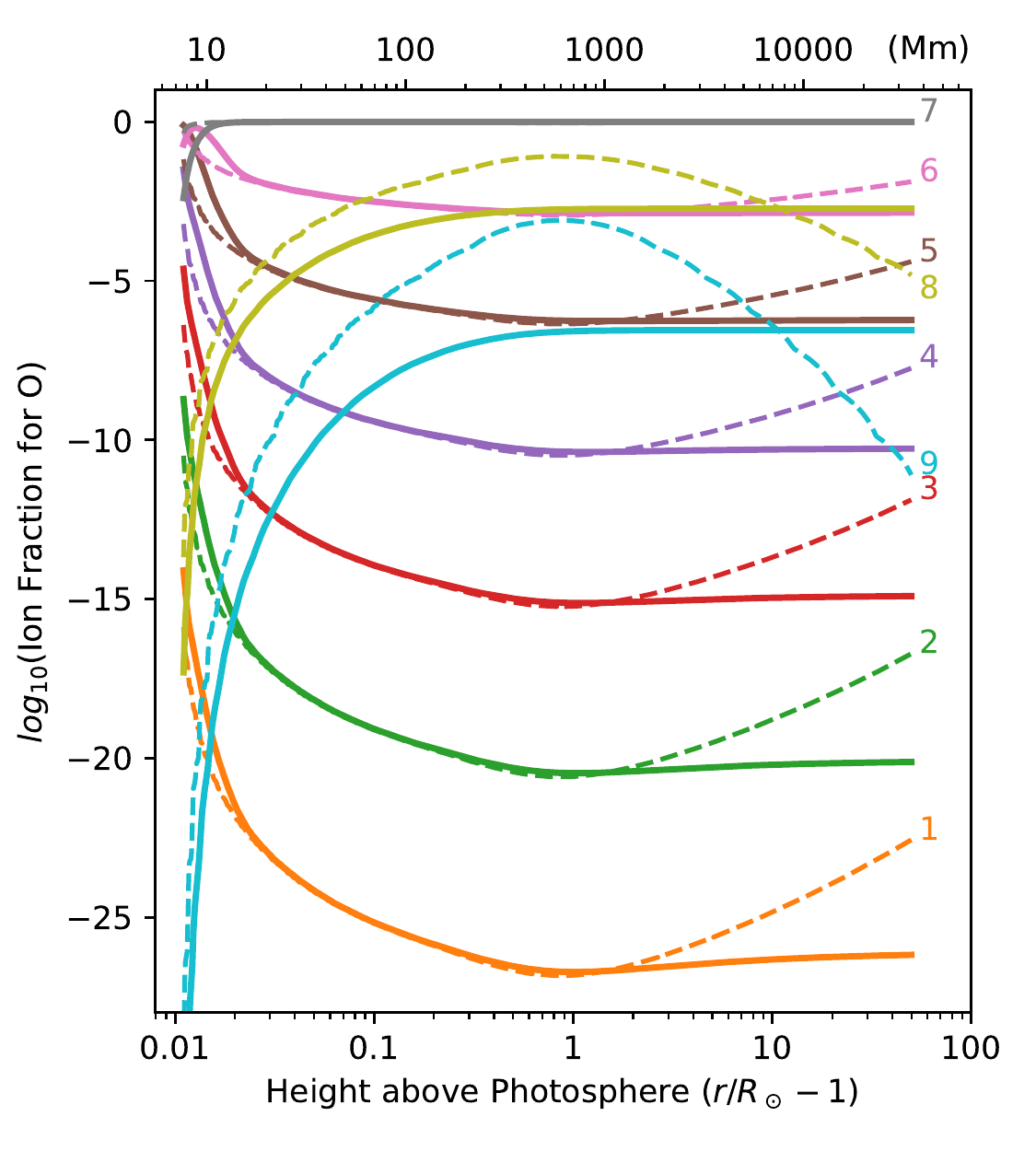}
	\caption{The ionization fractions $n_i/n_Z$ of each of the ions of oxygen. Dashed lines show the equilibrium calculation, and solid lines show numerical solutions to Equation (\ref{eq:ionization}).}
	\label{fig:chargestatecalc}
	%    	\end{center}
	\end{figure}

\begin{figure}[ht!]
		\begin{center}
			\includegraphics[width=\linewidthb]{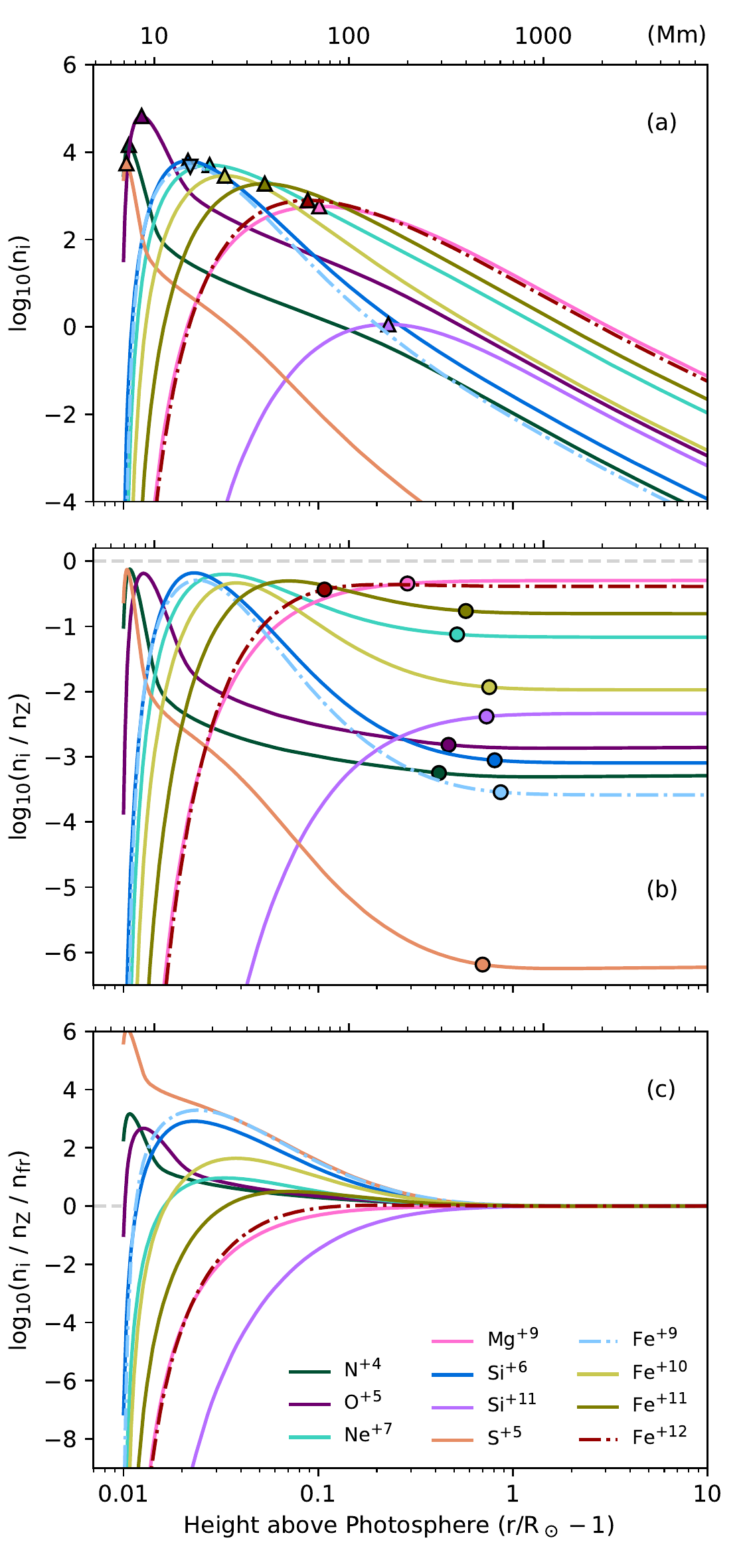}
			\caption{Ionization fractions for the ions from Table \ref{tab:lineChoices}. (a) Absolute number density of each species $n_i$ (in units of cm$^{-3}$), with triangles showing the peak density. (b) Charge state fractions shown as $n_i/n_Z$. Circles show the freezing-in height. (c) Fractions normalized to their frozen-in values at $r=10R_\odot$.}
			\label{fig:chargestatecalc2}
		\end{center}
	\end{figure}

    Expanding Equation (\ref{eq:rawIonization}) gives
    
    \begin{equation} \label{eq:ionization}
        \frac{\partial n_i}{\partial r} = \frac{n_e I_i}{u}  - \frac{2n_i}{r} - \frac{n_i}{u} \frac{du}{dr} - \frac{n_i}{f}\frac{df}{dr},
    \end{equation}
    which we solve numerically. For a stiff set of equations like these,we used the function ``\verb|solve_ivp|" from the Scipy ``integrate" package, which utilizes an implicit Runge-Kutta method of the Radau IIA family of order 5 \citep{Hairer1981, Oliphant2007}. The charge states were thus determined using solar wind, density, and temperature data from ZEPHYR, with initial conditions provided by the equilibrium calculation at $r = 1.0015 R_{\odot}$ (deep in the chromosphere). Representative results can be seen in Figures \ref{fig:chargestatecalc} and \ref{fig:chargestatecalc2}. 
    
	In Figure \ref{fig:chargestatecalc}, the charge states for all ions of oxygen are displayed. Notice that even within a single element, different ion populations can have very different behavior as a function of height. There are two main types of behavior, with ions either increasing in number density with height or decreasing with height. 
	
    For the decreasing ions, there are three distinct regions of interest. At the lowest heights, from about $r=1.01R_\odot$ to $r=1.03 R_\odot$, the ions are not in collisional equilibrium. As the temperature rises with height, the advective term $\partial n_i/\partial r$ balances first $n_e n_{i-1} C_{i-1}/u$ (as excess ions are collisionally excited into the state $i$) and then $n_e n_{i} C_{i}/u$ (as excess ions are subsequently ionized up into the next state $i+1$ and equilibrium is restored). This is similar to the so-called ``cold-effect" identified by \cite{Landi2012, Landi2012c}. By about $r = 1.03 R_\odot$, these ions have reached equilibrium, with collisional terms from Equation (\ref{eq:ionTerms}) balancing each other, while the density gradient $(\partial n_i/\partial r)$ is flatter now that the corona is high temperature with a large scale height. This equilibrium lasts until each ion's freezing-in radius $R_{fr}$, at which point the collisional terms have dropped off and the equation becomes a balance between the advective terms $\partial n_i/\partial r$ and $2 n_i/r$. The flattening of the non-equilibrium density curves in Figure \ref{fig:chargestatecalc} demonstrates this freezing-in behavior. 
    
    For the increasing ions, there are also three regions. The lower disequilibrium  still exists, primarily balancing $\partial n_i/\partial r$ and $n_e n_{i-1} C_{i-1}/u$, and it extends to larger heights than for decreasing ions (as high as $r=1.4R_\odot$ in some cases).  Then the advective terms become dominant again, first with $\partial n_i/\partial r$ and $(n_i/u) du/dr$ dominating, then with $\partial n_i/\partial r$ and $2 n_i/r$ balancing above the freezing heights $R_{fr}$, as before.

    In Figure \ref{fig:chargestatecalc2}(a), the non-equilibrium density profiles for all of the ions with spectral lines modeled in this paper (see Table \ref{tab:lineChoices}) are shown as a function of height. The triangle markers indicate the height of maximum absolute density for each ion, which we call the ion's peak radius $R_{p}$ and discuss further in Section \ref{sec:thermalResults}. Figure \ref{fig:chargestatecalc2}(b) normalizes these curves by the the total number density $n_Z(r)$ of each element, which constructs the charge state fraction. The freezing-in behavior is most evident in this panel, and the freezing-in radius $R_{fr}$ for each ion is marked with a circle. Figure \ref{fig:chargestatecalc2} (c) further normalizes each curve to their values at $r=10R_\odot$, which helps to demonstrate the different types of behavior an ion can display in the lower corona. For example, S$^{+5}$ has a much higher density at low heights than its frozen-in value, while Si$^{+11}$ has a much lower density than its frozen-in value. The impact of these different behaviors on off-limb line emission significant, and is explored in Section \ref{sec:thermalResults}.
    
%    It is true that the FIP effect is taken into account because we are using the coronal abundances from Chianti. 

\subsection{Coronal Hole Geometry: Superradial Expansion} \label{sec:superradialExpansion}

    Coronal holes exist over the poles of the Sun, caused by the concentration of a single magnetic polarity in those regions. Only a small fraction of the solid angle of the solar surface is composed of such coronal holes, which have open fields that reach out into the solar system. In contrast, over much of the solar cycle the equatorial region of the Sun is covered in closed-field regions.  Therefore, the solid angle subtended by a polar coronal hole is assumed to increase as it expands high into the corona, from roughly $0.5$ steradians at the photosphere up to until it eventually subtends $2\pi$ steradians at infinity. This superradial expansion has been measured by, e.g., \cite{Munro1977}, \cite{Guhathakurta1994}, and \cite{Deforest12001}.
    
    To model the superradial expansion of the coronal hole, we assume that the expansion happens self-similarly everywhere, in an idealized axisymmetric polar cap \citep[see also][]{Cranmer1999}. The radial variation in the total area of a circular coronal hole can be described by
    \begin{equation} \label{eq:areaF}
    	A(r) = A(R_\odot)\left(\frac{r}{R_\odot}\right)^2f(r) ,
    \end{equation}
    where $f(r)$ represents the superradial expansion of the flux tubes. We solve for $f(r)$ using the ZEPHYR model's input magnetic field strength, recalling that in a magnetic flux tube, $A(r)\propto |B(r)|^{-1}$. This factor is then used to determine the direction of the magnetic field $\hat{B}$ everywhere in the corona. It is also used in the calculation of the ion densities $n_i$ as a function of height (see Equation \ref{eq:rawIonization}).

    \begin{figure}[ht!]
	\begin{center} 
		\includegraphics[width=\linewidthb]{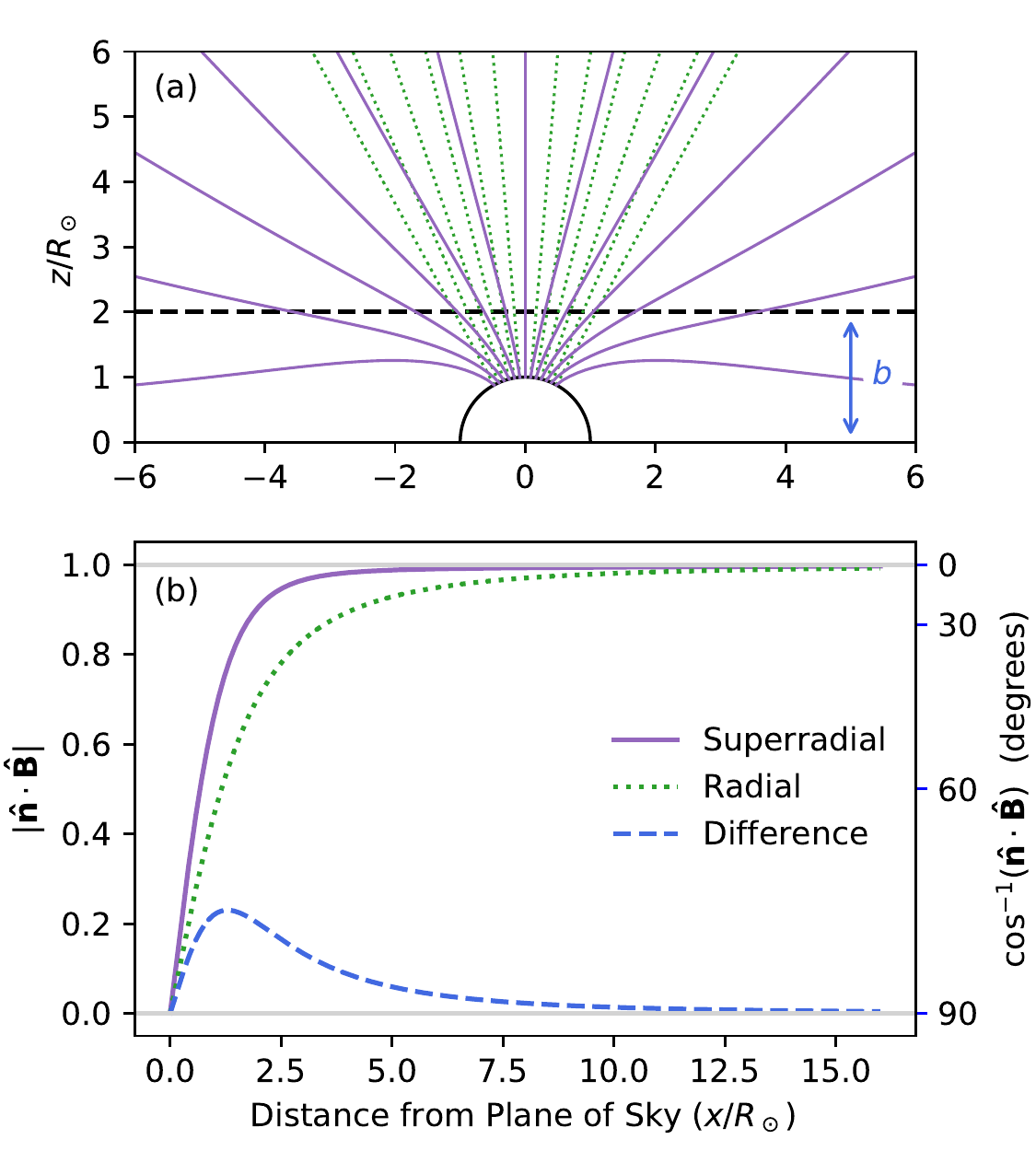}
		\caption{(a) Radial (green) and superradial (purple) expansion of magnetic field lines in the solar corona, with a sample LOS at impact parameter $b=2R_\odot$ (shown as a dashed line). (b) LOS dependence of the dot product of the LOS and the magnetic field $|\mathbf{\hat{n}}\cdot\mathbf{\hat{B}}|$, as well as the difference between the two cases. The right scale bar shows this as degrees above horizontal.}
		\label{fig:superRadial}
	\end{center}
\end{figure}
    
    Figure \ref{fig:superRadial}(a) shows the derived magnetic geometry in contrast to radial field lines. The magnetic field lines are deflected towards the equator in the superradial case, and because the solar wind is constrained to move along these magnetic field lines, this affects the direction of the solar wind flow $\mathbf{\hat{u}} = \mathbf{\hat{B}}$ as well. Figure \ref{fig:superRadial}(b) shows the dot product of the LOS direction $\mathbf{\hat{n}}$ and the magnetic field $\mathbf{\hat{B}}$ for a LOS with an impact parameter of $b=2R_\odot$. The difference between the cases is on the order of 20\% out to $2.5 R_\odot$, and 10\% out to 7 $R_\odot$. This extra deflection of the magnetic fields, especially near the plane of the sky (POS), allows the solar wind to broaden the spectral lines more than might be anticipated. This is examined in detail in Section \ref{sec:windResults}.

%\startlongtable
%	\centerwidetable
\movetableright=-1in
\begin{table*}[ht!] 

	\begin{center}
	
	\caption{Simulated Coronal Emission Lines \label{tab:lineChoices} }
	
    \begin{tabular}{lcccccccr}
	\hline
	Ion     & $\lambda_0$ & $T_{eq}$ & $q(T_{eq})$ & $E_1$ & $z_p$ & $z_{fr}$ & $z_p$ & $z_{fr}$ \\
	& ({\AA})  & $\log_{10}$(K) &  $\log_{10}$(cm$^3$ s$^{-1}$)  &      & {($R_{\odot}$)} & {($R_{\odot}$)} & {(Mm)} & {(Mm)}\\
	\hline
	
	\ion{N}{5}      &1238.82        &5.17    &--7.97        &1/2    &0.011  &0.417  &7.4    &290.6 \\
	\ion{O}{6}      &1031.91        &5.36    &--8.10        &1/2    &0.012  &0.467  &8.6    &325.4 \\
	\ion{O}{6}      &1037.61        &5.36    &--8.40        &0      &0.012  &0.467  &8.6    &325.4 \\
	\ion{Ne}{8}     &770.43 &5.69    &--8.35        &1/2    &0.028  &0.517  &19.3   &360.1 \\
	\ion{Mg}{10}    &624.97 &5.95    &--8.87        &0      &0.101  &0.287  &70.5   &200.1 \\
	\ion{Si}{7}     &275.36 &5.62    &--9.25        &7/20   &0.021  &0.807  &14.9   &561.6 \\
	\ion{Si}{12}    &499.41 &6.18    &--8.78        &1/2    &0.229  &0.732  &159.7  &509.9 \\
	\ion{S}{6}      &933.38 &5.12    &--7.87        &1/2    &0.010  &0.699  &7.2    &486.9 \\
	\ion{Fe}{10}    &184.54 &5.68    &--9.36        &0      &0.022  &0.866  &15.3   &603.2 \\
	\ion{Fe}{11}    &188.22 &5.76    &--8.99        &7/20   &0.033  &0.756  &23.1   &526.6 \\
	\ion{Fe}{12}    &195.12 &5.84    &--8.70        &7/25   &0.053  &0.574  &37.0   &399.4 \\
	\ion{Fe}{13}    &202.04 &5.91    &--8.55        &1      &0.088  &0.108  &61.5   &75.02 \\
	
	\end{tabular}

	\end{center}
	\ \\
	\ \\
\end{table*}

%    

%%%%%%%%%%%%%%%%%%%%%%%%%%%%%%%%%%%%%%%%%%%%%%%%%%%%%%%%%%%%%%%%%%%%%%%%%%%%%%%%%%%%%%%%%%%%%%%%%%%%%%%%%%%%%%%%%%%%%%%%%%%%%%%%%%%%%%%%%%%%%%%%%%
\section{Radiative Physics} \label{sec:radiative}

Here we describe the spectral lines we choose to model (Section \ref{sec:lineChoice}), our model for generating the emissivity in the extended corona and radiative transfer(Section \ref{sec:spectralLines}), the methods by which we reduce and parameterize the simulated observations (Section \ref{sec:understandingLines}), and some ways we can interpret and verify the validity of that reduction procedure (Section \ref{sec:validatingObservations}).

\subsection{Choice of Lines} \label{sec:lineChoice}
The lines we have chosen to examine can be seen in Table \ref{tab:lineChoices}. Included in the table are the rest wavelength $\lambda_0$, the equilibrium formation temperature $T_{eq}$ (the electron temperature at which that ion's equilibrium charge fraction is maximized), the collision strength at that temperature $q(T_{eq})$, and the scattering parameter $E_1$. We also calculate the heliocentric freeze-in radius $R_{fr}$ and the height of maximum absolute ion number density, which we call the peak radius $R_{p}$. We report these as the freeze-in height above the photosphere 
$z_{fr}=R_{fr}-1R_{\odot}$, and the peak height above the photosphere $z_{p}=R_{p}-1R_{\odot}$.

We selected these lines because their widths have been examined extensively in the context of off-limb coronal spectroscopy \citep[see, e.g.,][]{Noci1987, Kohl1997,Banerjee1998a, Landi2012b, Bemporad2012, Hahn2013, DelZanna2019}, and also because they represent a wide range of freezing-in heights, peak radii, and collisional/radiative intensity ratios. There are also some lines (e.g., the O VI doublet), where the interpretation of some measurements is less straightforward \citep[see, e.g.,][]{Tu1998, Kohl1999, Esser1999, Kohl2006, Cranmer2008a}. 

%[A major task I have is to go through all these references, as well as the ones in Section 4.1, and figure out what narrative I want to use and which papers support that narrative and where each one fits best. I also intend to digitize some relevant plots from these references and overplot them on some of my plots. This section and 4.1 will change a good bit]

%\cite{Esser1999} is the one about the hot OVI lines, also looks at exactly this stuff. \cite{Kohl1999} also has measurements of 1037, shows a v1/e that is flat up to 1.8 or so. \cite{Cranmer2008a} also studies the line widths of the OVI lines. Make some kind of statment here about that, or just move it to the introduction.

%%%%%%%%%%%%%%%%%%%%%%%%
\subsection{Spectral Line Formation} \label{sec:spectralLines}
	In this work we generate synthetic spectral lines by modeling the two main processes by which light is emitted by heavy ions in coronal plasma: collisional excitation, which dominates the line emission at low heights where collisions are frequent, and resonant scattering, which becomes more important in the higher regions of the atmosphere as collisions become rare. 
		
	For each synthetic observation, a LOS is defined through the corona as the $x$-axis of a Cartesian coordinate system, with $x=0$ (the POS) defined as directly above the solar north pole, $y=0$, and $z=b$ (i.e. a straight line going directly from the Earth over the pole of the Sun at a given distance $b$). The LOS extends into the foreground and background to a distance $s = S(b)$, described below. The plasma properties at each point along the LOS are then determined by interpolating the input data described in Section \ref{sec:backgroundPlasma}. These parameters are used to determine the local spectral emissivity
	\begin{equation}
	j(x,\nu) = j_c(x,\nu)+j_r(x,\nu), 
	\end{equation} 
	as described in the following sections.	
		
	The regions of the corona we study here exist high above the photosphere and are rarefied enough to be optically thin, which greatly simplifies the solutions of the equation of radiative transfer \citep[see, e.g.,][]{Withbroe1982, Olsen1994, Cranmer1999, Kohl2006}. This allows us to simply integrate the emissivity along the LOS to give the specific intensity
	\begin{equation}
	I(\nu) = \int_{-s}^{s}dx\ j(x,\nu), 
	\end{equation}
 	(i.e., a spectral line) which is analyzed according to the procedure in Section \ref{sec:understandingLines}. One of the advantages to the 
 	forward 
 	modeling approach is that we can also examine the total emissivity along the LOS
 	\begin{equation}
 	J(x) = \int d\nu\ j(x,\nu).
 	\end{equation}
 	
 	GHOSTS uses variable resolution along the LOS $x$-axis, choosing the smallest grid spacing $\Delta x$ from the following rules: $\Delta x=0.004 
 	R_\odot$ for $r \leq 2 R_\odot$, $\Delta x=0.02 R_\odot$ for $|x| \leq 5 R_\odot$, and $\Delta x=0.2 R_\odot$ for $|x|> 5R_\odot$. 
 	
	We performed a study to determine $S(b)$, the minimum extent in and out of the POS that must be simulated to achieve accurate results. A lower value significantly decreases computation time, but a value that is too low will begin to truncate and modify the simulation, especially at the top of the domain where the POS is less dominant. For measurements taken up to a height $b$, there should be a critical value $S(b)$ above which the results do not change, with lower values altering the results. For this model, we found that when looking up to $b=3 R_\odot$, the results are stable with $S(3) \geq 20 R_\odot$, with lower values altering the results by up to 10\%. When looking up to $b= 6 R_\odot$, $S(6) \geq 35 R_\odot$ must be simulated, and when looking up to $b=11 R_\odot$, $S(11) \geq 50 R_\odot$ is required. For this work, we choose to use $S(b) = 75 R_\odot$ for all heights $b$ to ensure consistent and valid result for our domain of $b=1.01$ to $11 R_\odot$.

%%%%%%%%%%%%
\subsubsection{Collisional Excitation} \label{sec:collisionalLines}
	
	Collisional excitation occurs in dense plasmas when a free electron collides with an ion, lending its kinetic energy to a bound electron momentarily before the energy is radiated away as a photon. This process occurs frequently, each photon getting a slight Doppler shift from the random dynamics of the collision. For ions and electrons with Maxwellian velocity distributions, following \cite{Withbroe1970}, this process can be modeled as a Gaussian spectral line being produced by each point along the LOS. The emissivity is given by
	\begin{equation}
	j_{c}(x,\nu)= \frac{h\nu_0}{4\pi} n_e n_i q(T_e) \Phi(\nu),
	\end{equation}
	where $h$ is Planck's constant, $\nu_0$ is the rest frequency of the line, $T_e$ is the local electron temperature from ZEPHYR, and $q(T_e)$ is the temperature-sensitive collision strength of the ion from CHIANTI. The shape of the spectral line is given by the line profile function $\Phi$:
	\begin{equation} \label{eq:lineProfile}
		\Phi (\nu) \, = \, \frac{1}{\Delta \nu \sqrt{\pi}} \exp \left[
		- \left( \frac{\nu - \nu_0 - \nu_{\rm los}}{\Delta\nu} 
		\right)^2 \right] \,\,\, ,
	\end{equation}
	where
	\begin{equation}
		\Delta \nu \, = \, \frac{v_{th}}{c} \nu_0
		\,\,\,\,\,\,\,\,\,\,\,
		\mbox{and}
		\,\,\,\,\,\,\,\,\,\,\,
		\nu_{los} \, = \, \frac{v_{los}}{c} \nu_0
		\label{eq:freqShift}
	\end{equation}
	are the thermal width and the Doppler-shift of the line, $v_{th} = \sqrt{2k_bT_i/m_i}$ is the thermal velocity of the ion, $v_{los} = 
	\mathbf{u}\cdot\enn$ is the component of the point's bulk velocity $\mathbf{u}$ that is projected into the LOS direction $\enn$ (with positive velocity towards the observer), $c$ is the speed of light, $m_i$ is the mass of the ion, and $T_i$ is the ion temperature. For most of this paper, $T_i = T_e$, but the effects of preferential ion heating are discussed in Section \ref{sec:boostResults}.

%%%%%%%%%%%%
\subsubsection{Resonant Scattering} \label{sec:resonantLines}
	When radial light from the solar photosphere $I_0(\nu')$ interacts with coronal ions, it can be scattered from it's incident direction $\enn'$ 
	into the LOS direction $\enn$. This produces a resonantly scattered component of the line, with an emissivity given by
	\begin{equation} \label{eq:rezemissivity}
	j_{r}(x,\nu) = \frac{h\nu_0}{4\pi} n_i B_{12} \int \frac{d\Omega'}{4\pi} \int d\nu' {\cal R}(\nu',\enn'; \nu, \enn)  I_0(\nu'),
	\end{equation}
	\noindent where $B_{12}$ is the Einstein absorption rate of the transition and the two integrals are taken over the distributions of incoming 
	photon directions and frequencies. See Appendix \ref{sec:studyingR} for a discussion of how the limits of the integral taken over $\nu'$ must be 
	chosen carefully to avoid truncating the results. For simplicity and speed, we approximate the integral over incident solid angle by assuming a single ray of light from the 
	center of the Sun $\enn'=\mathbf{\hat{r}}$, using a dilution factor given by
	\begin{equation}
	W(r) = \int \frac{d\Omega'}{4\pi} = \frac{1}{2} \left( 1 - \sqrt{1- (1/r)^2} \right).
	\end{equation}	
	to model the radial decrease in incident intensity. Note that some applications (i.e., modeling polarized light or the effects of highly anisotropic ion velocity distributions) require the full solid-angle integral to be calculated. 
	
	We use the Case I photon redistribution function ${\cal R}$ as discussed in detail by \cite{Withbroe1982} and \cite{Cranmer1998}, given by
	\begin{equation}
		{\cal R}(\nu', \enn'; \nu, \enn) = \frac{g(\theta)}{\pi\beta(\Delta\nu)^2} \exp\left[-\zeta'^2 - \left(\frac{\zeta-\alpha\zeta'}{\beta}\right)^2\right],
		\label{eq:resonantEmiss}
	\end{equation}
	where
	\begin{equation}
	\zeta = \frac{\nu-\nu_0}{\Delta\nu} - \frac{v_{los}}{v_{th}},
	\,\,\,\,\, \text{and}\,\,\,\,\,\,
	\zeta' = \frac{\nu'-\nu_0}{\Delta\nu} - \frac{v'}{v_{th}}.
	\label{eq:dimensionlessShift}
	\end{equation}
	The scattering ion has a relative velocity with both the photospheric emission $v' = \mathbf{u}\cdot\enn'$ and the observer $v_{los} = 	\mathbf{u}\cdot\enn$, which 
	causes complex Doppler shifting. The scattered emissivity is also modulated as a function of the scattering angle $\theta = \cos^{-1}(\enn' \cdot 
	\enn)$, with $\alpha=\cos\theta$ and $\beta=\sin\theta$. The phase function is
	\begin{equation}
	g (\theta) = \left( 1 - \frac{E_1}{4} \right) + \frac{3E_1}{4} \alpha^2,
	\end{equation}
	for the simple case of an unpolarized incoming intensity. We use the \cite{Chandrasekhar1960} definition of $E_1$, which gives isotropic 
	scattering for $E_1 = 0$ and Thomson/Rayleigh scattering for $E_1 = 1$, and we provide the value of $E_1$ for each line in Table 
	\ref{tab:lineChoices}. 
	
	The photospheric intensity spectrum $I_0(\nu')$ was assembled from two sources. For $\lambda > 670$~{\AA}, we used the quiet-Sun spectral atlas obtained by the Solar Ultraviolet Measurements of Emitted Radiation (SUMER) spectrometer aboard {\em SOHO} \citep{Curdt1997}. For $\lambda < 670$~{\AA}, we used EUV irradiance data provided online by T.\  Woods from a series of rocket calibration flights of the Multiple Extreme-ultraviolet Grating Spectrographs (MEGS) between 2008 and 2013 \citep[see, e.g.,][]{Hock2012}. The SUMER data came from a less active phase of the solar cycle than the MEGS data, so the intensities of the latter were divided by a constant factor of 4.0 to produce a single consistent spectrum. This factor was determined by cross-calibrating the two spectra over the overlapping wavelength range of 670--680 {\AA}.
	
	The requirements for resolution and spectral range for the redistribution computation are stringent, and are discussed in more detail in Appendix \ref{sec:studyingR}. In summary, the significance of Doppler pumping and dimming on the measured line widths means that it is important to specify a broad enough portion of the incident spectrum to avoid truncating and invalidating the results. Additionally, the nature of the sparse diagonal matrix ${\cal R}$ necessitates high resolution calculations to avoid aliasing.

%%%%%%%%%%%%%%%%%%%%%%%%

\subsection{Reducing Spectral Line Observations} \label{sec:understandingLines}
	
	\newcommand{\Tau}{\mathcal{T}} % This is the symbol for the derived line temperature
	\newcommand{\VV}{\mathcal{V}} % This is the symbol for the derived line velocity
	\newcommand{\ww}{\eta}  % This is the symbol for the weighting functions
	\newcommand{\wt}{\langle T \rangle}
	\newcommand{\wc}{\langle T_C \rangle}
	\newcommand{\wrr}{\langle T_R \rangle}
	\newcommand{\uu}{\langle U \rangle}
	\newcommand{\uc}{\langle U_C \rangle}
	\newcommand{\vu}{\langle {\cal V} \rangle}
	\newcommand{\urr}{\langle U_R \rangle}
	\newcommand{\xiw}{\xi_W}

	It is advantageous to reduce the full specific intensity profile of the spectral line $I(\nu)$ to a small set of parameters which can be easily compared between different lines. Due to the finite resolution and count-rate statistics in most coronal spectrometers, it is a common procedure to analyze spectral lines by fitting them with a simple Gaussian, even though this could in principle throw out some detailed physical information. The scipy routine ``\verb|curve_fit|" was used to fit a Gaussian profile to each simulated spectral line, returning fit amplitudes $A$, centroids $\nu_0$ and $1/e$ half-widths $\Delta \nu$ for each line. In this work, we focus primarily on understanding the spectral width $\Delta \nu$.
	
	Because LOS-projected bulk velocity Doppler-shift's the emitted light and broadens the line, it is natural to express the $1/e$ spectral width $\Delta \nu$ as a velocity, such that
	\begin{equation}\label{eq:widthConvert}
	\VV \,\, \equiv \,\, \frac{\Delta\nu}{\nu_0}c.
	\end{equation}
	This quantity will be examined extensively in the following sections, and represents the full measured line width. The interpretation of this width requires some care, however, as $\VV$ can be understood as being composed of a thermal and a nonthermal component that are blended together. Even in the absence of any macroscopic motions along the LOS, there will be a width $V_{th}$ due to the random thermal motions of the emitting particles, and any bulk flow velocities along the LOS cause additional non-thermal broadening $\xi$. If the thermal and nonthermal components are assumed to be Gaussian in form, the total measured width of the spectral line can be expressed as
	\begin{equation}\label{eq:decompose}
	\VV^2 \,\, = \,\, V^2_{th} + \xi^2 \,\, = \,\, \frac{2k_b\Tau}{m_i} + \xi^2 \, ,
	\end{equation}
	where the thermal term contains the temperature-like quantity $\Tau$, representing the observed temperature.
		
	Unfortunately, one can only measure the total line width $\VV$, and there does not seem to be a model-independent way to know how much of that width comes from thermal effects and how much comes from each of the different types of bulk flow (solar wind, Alfv\'en waves, etc.). If one were known in some other way, however, then the other could be determined. One way to gain some insight with minimal assumptions is to look at two limits: the so-called kinetic temperature $\Tau_k$ can be formed we assume no nonthermal broadening ($\xi = 0$), and $\xi_{\text{max}}$, where we assume $\Tau=0$. This is an excellent way to provide relatively strong upper and lower bounds on the range of possible conditions \citep[see, e.g.,][]{Tu1998}, but it cannot give an exact answer for either quantity. 
	
	A more sophisticated version of this approach involves using a model to make a better choice for the secondary quantity: what is $\Tau$ if we subtract a modeled value for $\xi$, or (more commonly) what is $\xi$ if we subtract a modeled or observationally determined thermal component $\Tau$? The choice of $\Tau$ makes a big difference in the result for $\xi$, so in Section \ref{sec:windResults} we examine two common choices to see which one performs better at recovering the radial variation of the input model.  In one case we use the target input value of $T_i$ in the POS from ZEPHYR, and in the other case use the weighted temperature $\wt$ (as defined in the next section) as a proxy for an observationally determined temperature. These are defined as 
	\begin{equation}\label{eq:reduceThermal}
	\xi_P \equiv \sqrt{\VV^2 - \frac{2k_bT_i}{m_i}}
	\end{equation} 	
	and
	\begin{equation}\label{eq:reduceThermal_wt}
	\xiw \equiv \sqrt{\VV^2 - \frac{2k_b\wt}{m_i}}.
	\end{equation} 		
	
	One way to try to disambiguate $\Tau$ and $\xi$ has been to recognize that $V_{th}$ depends on the mass of the emitting ion, but most models for $\xi$ do not. This could allow multi-ion observations to attempt to tease out a temperature component \citep[see, e.g.,][]{Seely1997, Moran2003}. This method is examined in Appendix \ref{sec:multiIon}, and we find that it does not seem to work reliably in the presence of solar wind to determine $\Tau$, but is able to retrieve $\xi$ to some degree.

\subsection{Validating Line Width Reductions} \label{sec:validatingObservations}

Once the measurements have been reduced to $\VV$, $\xi$, and $\Tau$, they still require careful analysis. It is important to keep in mind that the observed LOS contains plasma of many different temperatures and densities, and light from the entire column is being summed to produce the observed spectral line. It is difficult to be certain which part of the LOS (if any!) is well-described by the derived parameters. One would like to believe that $\Tau$ is something like an emissivity-weighted average of the local ion temperatures $T_i$, and that $\xi$ provides an emissivity-weighted measure of the LOS bulk velocity components. 

To validate these assumptions, we leverage our forward model to construct emissivity-weighted average quantities directly from the LOS plasma parameters, which can be used as comparisons to the reduced line-width measurements. If they match, then we can say that we understand what the derived quantities are measuring. Thus, our task is to find new quantities $\vu \approx \VV,  \wt \approx \Tau,$ and $ \uu \approx \xi$. We use $U$ because in this work we deal only with bulk flows and not waves.

The expected behaviors of different kinds of velocity fields can be examined by constructing a generalized second-moment frequency width, 
\begin{equation} \label{eq:second-moment}
\langle \delta \nu^2 \rangle \, = \, \frac{\int d\nu \,\, 2\nu^2 \, \int dx \, J(x) \, \Phi(\nu,x)} {\int d\nu \, \int dx \, J(x) \, \Phi(\nu,x)} ,
\end{equation} 
and note the factor of two in the numerator. Without it, the numerator would give the straightforward variance moment of $\Phi(\nu,x)$. The square root of $\langle \delta \nu^2 \rangle$ is the standard deviation, not the $1/e$ half-width that we use elsewhere in this paper. Recall that $J(x)$ is the total emissivity at $x$.

First let us construct a match for $\Tau$. For a single thermal Gaussian,
\begin{equation} 
\Phi (\nu, x, a) \, = \, \frac{1}{a \sqrt{\pi}} \exp \left[ - \frac{(\nu - \nu_0)^2}{a^2} \right] 
\end{equation}
at each point along the line of sight. Ignoring bulk velocities, the integrals in Equation (\ref{eq:second-moment}) give 
\begin{equation} 
\langle \delta \nu^2 \rangle \, = \, \frac{\int dx \, J(x) \, a^2}{\int dx \, J(x)}\,\, .
\end{equation} 
We use this form to define the modeled emissivity-weighted thermal width
\begin{equation} \label{eq:losAverageT}
\wt \equiv \dfrac{m_i}{2k_b}\langle \delta v_{th}^2 \rangle \, = \dfrac{m_i}{2k_b}\dfrac{\int^s_{-s}dx\ J(x)\ v_{th}^2 }{ \int^s_{-s}dx\ J(x)},
\end{equation} 
in temperature units. 

Now we will add in the effects of bulk velocity. If the nonthermal velocity field takes the form of a coherent Doppler shift due to a bulk LOS velocity (i.e. the solar wind), the idealized line profile function becomes 
\begin{equation} 
\Phi (\nu, x, a, b) \, = \, \frac{1}{a \sqrt{\pi}} \exp \left[ - \left( \frac{\nu -\nu_0 - b}{a} \right)^2 \right] 
\end{equation} 
(compare to Equation (\ref{eq:lineProfile})), and the integrals in Equation (\ref{eq:second-moment}) give 
\begin{equation} 
\langle \delta \nu^2 \rangle \, = \, \frac{\int dx \, J(x) \, (a^2 + 2b^2)}{\int dx \, J(x)}\,\, .
\end{equation}
We use this to define the modeled emissivity-weighted LOS solar wind speed
\begin{equation} \label{eq:losAverageU} 
\uu \equiv \sqrt{\dfrac{\int^s_{-s}dx\,\, J(x)\ (2v_{los}^2)}{ \int^s_{-s}dx\ J(x)}},
\end{equation} as well as the modeled emissivity-weighted total line broadening
\begin{equation} \label{eq:losAverageV} 
\vu \equiv \sqrt{\dfrac{\int^s_{-s}dx\,\, J(x)\ (v_{th}^2 + 2v_{los}^2)}{ \int^s_{-s}dx\ J(x)}}.
\end{equation}

Throughout this work, $\wt$, $\uu$, and $\vu$ will be described as ``modeled quantities" because they are determined directly from the detailed LOS plasma parameter information, which is more similar to the information available to a modeler. On the other hand, $\Tau$, $\xi$, and $\VV$ will be described as ``measured" or ``observed" quantities, as they are determined from (simulated) spectral line profile intensities, which is similar to the type of LOS-integrated information available to an observer.

Ion independent versions of these emissivity-weighted quantities can be computed by replacing $J$ with another parameter which has the same type of dependencies. We use $\tilde J_C=\rho^2$ as a proxy for collisional emissivity $J_C(x)$ and $\tilde J_R=\rho W$ as a proxy for resonant emissivity $J_R(x)$ to create the density-weighted quantities $\wrr$ and $\wc$. Note that both $J_C$ and $J_R$ behave as $\propto r^{-4}$ at large heliocentric distances, but their behaviors in the low corona are quite different. There does not seem to be an ion-independent way to build $\tilde J=J_C+J_R$, so these density-weighted curves are not precisely similar to the emissivity-weighted curves, but they serve as useful bounds that are easy to calculate and compare with. 

To briefly address wave phenomena, let us examine one more case. If there exists a nonthermal velocity field that is randomly incoherent (i.e., with multiple uncorrelated parcels along the LOS), its line profile function could be represented by a thermal Gaussian convolved with another Gaussian: 
\begin{equation}
\Phi (\nu, x) \, = \Phi (\nu, x, a) * \Phi (\nu, x, b)\,\, .
\end{equation}
For constant values of the widths $a$ and $b$, this is equivalent to a single broader Gaussian, 
\begin{equation} 
\Phi (\nu, x, a, b) \, = \, \frac{1}{\sqrt{\pi (a^2 + b^2)}} \exp \left[ - \frac{(\nu-\nu_0)^2}{a^2 + b^2} \right]
\end{equation} 
and the effective width is given by 
\begin{equation} 
\langle \delta \nu^2 \rangle \, = \, \frac{\int dx \, J(x) \, (a^2 + b^2)}{\int dx \, J(x)} \,\, . 
\end{equation}
This explains the origin of the traditional way of combining thermal and nonthermal velocities via quadrature, as described by Equation (\ref{eq:decompose}). The nonthermal velocity is typically thought to be caused primarily due to Alfv\'en waves. Due to finite instrument integration times, the time-varying Alfv\'en waves crossing the field of view act as local microturbulent broadening, rather than a coherent Doppler shift, and it acts very similarly to an increased local temperature. Because it is often assumed that only a small region of the LOS contributes meaningfully to the observation, $a$ and $b$ are considered to be constant along the significant portions of the LOS, and determining their values is the goal of the observation.  We are not treating Alfv\'en waves at this time, so this form of the width is not appropriate. When waves are added to the model, however, we will have to consider a new version of $\vu$ that combines these two types of non-thermal effects. It is also worth noting that this paradigm ignores the fact that there are LOS effects, and that the values of $a$ and $b$ cannot be considered constant over the LOS, as we discuss in the following section.

%%%%%%%%%%%%%%%%%%%%%%%%%%%%%%%%%%%%%%%%%%%%%%%%%%%%%%%%%%%%%%%%%%%%%%%%%%%%%%%%%%%%%%%%%%%%%%%%%%%%%%%%%%%%%%%%%%%%%%%%%%%%%%%%%%%%%%%%%%%%%%%%%%%%%%%%%%%%%%%%%%%%%%%%%%%%%%%
%%%%%%%%%%%%%%%%%%%%%%%%%%%%%%%%%%%%%%%%%%%%%%%%%%%%%%%%%%%%%%%%        Flow Free Results Section         %%%%%%%%%%%%%%%%%%%%%%%%%%%%%%%%%%%%%%%%%%%%%%%%%%%%%%%%%%%%%%%%%%%%%
%%%%%%%%%%%%%%%%%%%%%%%%%%%%%%%%%%%%%%%%%%%%%%%%%%%%%%%%%%%%%%%%%%%%%%%%%%%%%%%%%%%%%%%%%%%%%%%%%%%%%%%%%%%%%%%%%%%%%%%%%%%%%%%%%%%%%%%%%%%%%%%%%%%%%%%%%%%%%%%%%%%%%%%%%%%%%%%
\section{Results} \label{sec:allResults}
Here we describe the simulations that were undertaken and what we found. Section \ref{sec:thermalResults} will examine a case with no macroscopic flow velocities, Section \ref{sec:windResults} will add in the influence of the solar wind, and Section \ref{sec:boostResults} will examine cases with preferential ion heating.

\subsection{Flow-Free Results} \label{sec:thermalResults} \label{sec:steadyResults}

%\begin{figure*}[ht!]
%	\includegraphics[width=\linewidth]{projFig.pdf}
%	\caption{(a) Line-fit temperature measurements $\Tau$ in flow-free ($B=0$) case. Triangles denote height of maximum ion number density $T_{eq}$. Dotted vertical line marks observation height $b$ of LOS in Figure \ref{fig:floorIntensity}. (b) Line fit velocity measurements $\VV$ in full-wind ($B=1$) case. (c),(d) Observations compared to radial value. (e),(f) Observations compared to the model equations.}
%	\label{fig:weighting} \label{fig:tempResults}
%\end{figure*}

\begin{figure}[ht!]
	\includegraphics[width=0.99\linewidth]{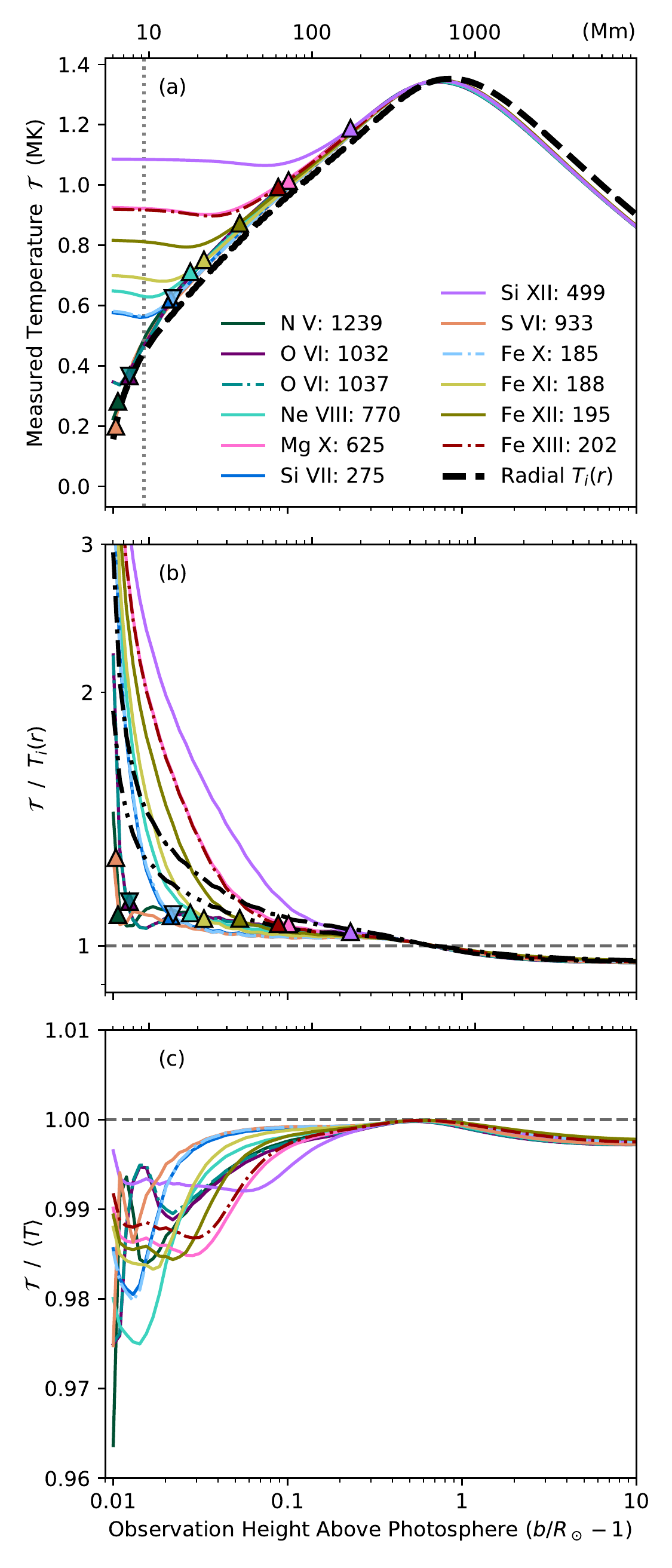}
	\caption{(a) Line-fit temperature measurements $\Tau$ in flow-free ($B=0$) case. Triangles denote height of maximum ion number density $T_{eq}$. Dotted vertical line marks the observation height $b$ of the LOS in Figure \ref{fig:floorIntensity}. Dash-dot and dash-double-dot curves show $\wrr$ and $\wc$, respectively. (b) Observations normalized to radial variation of $\mathbf{T_i(r)}$. (c) Observations compared to the model $\wt$.}
	\label{fig:weighting} \label{fig:tempResults}
\end{figure}

We start by examining a case in which the effect of the solar wind outflow on line broadening is ignored. During the calculation of the emissivity $j(x,\nu)$, we set the value of the bulk velocity $\mathbf{u}=0$ everywhere, such that line widths are only dependent on $T_i(x)$ and $n_i(x)$. We retain the frozen-in ionization balance discussed in Section \ref{sec:eqChargeStates}, however, despite its dependence on the solar wind outflow. This results in a slightly non-self-consistent situation, but we find it illustrative to first examine the thermal widths, then add in the effects of nonthermal broadening from the solar wind in the next section. A major goal of this kind of observation is to determine the value of $T_i(r)$ from the observed data, and we will examine whether this is feasible in cases with negligible wind broadening.

The 12 emission lines listed in Table \ref{tab:lineChoices} were synthesized for lines of sight over the pole with a range of impact parameters between $b = 1.01 R_\odot$ and $b = 11R_\odot$. Then $\VV$ was determined for each line using Equation (\ref{eq:widthConvert}). Because $\mathbf{u}=0$ everywhere, we can assume $\xi$ to be zero as well. This allows a straightforward conversion of the measured line width $\VV$ to $\Tau$ using Equation (\ref{eq:decompose}). The solid curves in Figure \ref{fig:weighting}(a) show $\Tau$ for each ion line as a function of observation height, and the dashed black line represents the input radial ion temperature $T_i(r)$ from ZEPHYR. Figure \ref{fig:weighting}(b) normalizes these curves to $T_i(r)$. Ion-independent curves $\wrr$ and $\wc$ are shown as the dash-dot and dash-double-dot curves, respectively. Note that while the behavior is similar to the curves for each ion, using the full $\rho$ density to weight the temperature does not closely agree with any of the simulated ion measurements in the lower corona. Figures \ref{fig:weighting}(a) and \ref{fig:weighting}(b) show good agreement between the simulated observables and the input model in the upper regions of the observation, but the lower regions are distorted. Assuming that $\Tau(b)\approx T_{i}(r)$ is reasonable for the top of the domain, but it is clearly not valid in the lower corona for many of the lines. 

On the other hand, Figure \ref{fig:weighting}(c) shows $\Tau$ compared with $\wt$, and it seems to be an accurate model to within a half of a percent above $b=1.1$, and within $2\%$ down to about $b=1.01$. In the absence of any bulk flows or waves, it appears that the width of a spectral line can provide an emissivity-weighted average of the LOS temperature as described by Equation (\ref{eq:losAverageT}). It is just important to keep in mind that $\Tau \approx \wt \neq T_i(r)$, especially below $b=1.1R_\odot$. 

\begin{figure}[ht!]
 	\includegraphics[width=\linewidth]{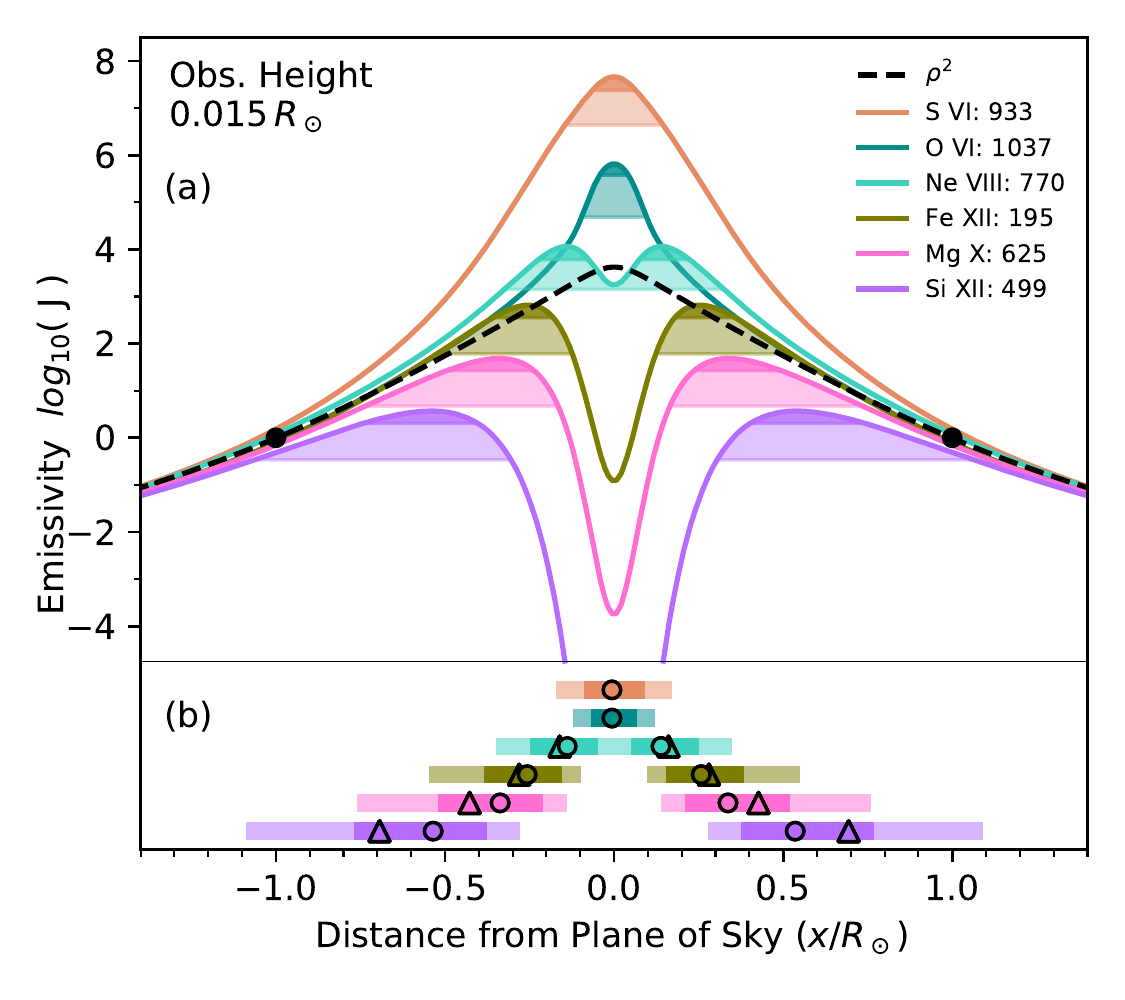}
 	\caption{
 		(a) LOS dependence of the total relative emissivity $J(x)$ for several ion lines for $b=1.015 R_\odot$. Curves are normalized first to their values at $x=3R_\odot$, then scaled to the value of $\rho^2$ at $x=1R_\odot$.
 		(b) A flattened representation of the dominant emissivity regions along the LOS. Circles mark the emissivity maxima, and triangles mark the projected peak radius.}
 	\label{fig:floorIntensity}
\end{figure}

In Section \ref{sec:eqChargeStates} we defined the peak height $R_p$ as the height at which a given ion's number density $n_i$ is maximized (marked as triangles in the figures). It is clear that $R_p$ correlates well with the deviation away from $T_i(r)$ in Figures \ref{fig:tempResults}(a) and \ref{fig:tempResults}(b). One could therefore also think of $R_p$ as a "plateau height." In this case, $\Tau$ approximates $T_i(r)$ to within a few percent for measurements taken above $b=R_p$, but shows little variation below that height, leading to observations that are off by a factor of 3 or more. Figure \ref{fig:chargestatecalc2} shows that in all of the modeled cases, the ion density drops off rapidly below $r=R_p$, and the region below that altitude does not contribute a significant amount of emissivity to the observation. This is illustrated in Figure \ref{fig:floorIntensity}(a), which shows the total emissivity $J(x)$ along a line of sight with a low impact parameter $b=1.015R_\odot$ (shown as a vertical dashed line in Figure \ref{fig:tempResults}(a)). For \ion{S}{6} 933 (the uppermost orange curve), whose emitting ion S$^{+5}$ has a very low $R_p=1.01 R_\odot$, the POS is dominant at this height and the observation matches the POS value. For \ion{Si}{12} 499 (the lowest violet curve), however, the POS density of Si$^{+11}$ is extremely rarefied this far below its peak radius of $R_p=1.229 R_\odot$, and the observed temperature exceeds the POS value by a factor of 3. This observation is actually ``measuring" the foreground and background plasma just below Si$^{+11}$'s peak radius. 

Figure \ref{fig:floorIntensity}(b) shows the regions that encompass the most dominant $68\%$ and $95\%$ portion of the LOS. Circles mark the location of maximum emissivity, while triangles show $x_{fl} = \sqrt{R_p^2 - b^2}$, the intersection of the peak radius with the LOS. The altitude of maximum emissivity is usually quite close to the projected peak radius, though the farther below it the measurement is taken, the worse that association becomes. Regardless, it is clear that while all these measurements were taken at the same observation height $b$, they are not sensitive to plasma conditions at the same heliocentric radius $r$. Any observation taken at an impact parameter $b < R_p$ will be dominated by the foreground and background plasma at $r \approx R_p$, which manifests as the measurement plateaus seen in Figure \ref{fig:tempResults}. Values of $R_p$ for each modeled line can be found in Table \ref{tab:lineChoices}, reported as height above the surface $z_p$. 

This type of plateaued observation does seem to exist in the literature. \cite{Landi2003} used SUMER to study quiet-Sun off-limb spectral lines from $b=1.00$ to $b= 1.35$, and their derived temperatures and non-thermal velocities seem roughly constant over that range. \cite{Andretta2012} used \textit{SOHO}/CDS to measure off-limb polar spectra, and they report approximately constant temperatures up to $b=1.2R_\odot$. \cite{Zanna2019} used \textit{Hinode}/EIS to measure the spectral widths of several lines of iron. They started with the assumption of constant temperature up to $r=1.5R_\odot$, and they concluded that there is no significant evidence for a variation of the excess (non-thermal) widths by more than 10 km s$^{-1}$ out to $b=1.3 R_\odot$. In contrast to these measurements, the modeled ZEPHYR $T_i(r)$ increases 120\% (from 0.5MK to 1.2MK) over the range $r=1.02 R_\odot$ (just above the transition region) to $r=1.3 R_\odot$, peaking at $T=1.35$ MK at $r=2 R_\odot$. It therefore seems likely that this floor effect is present in these observations. This is our first piece of evidence that spectral observing just above the limb of the Sun might falsely give the impression that these quantities are constant with height. We predict that multi-ion measurements made in higher regions of the corona will reveal ion-dependent $R_p$, which would be evidence that the floor effect is occurring.

\begin{figure*}[]
	\includegraphics[width=\linewidthb]{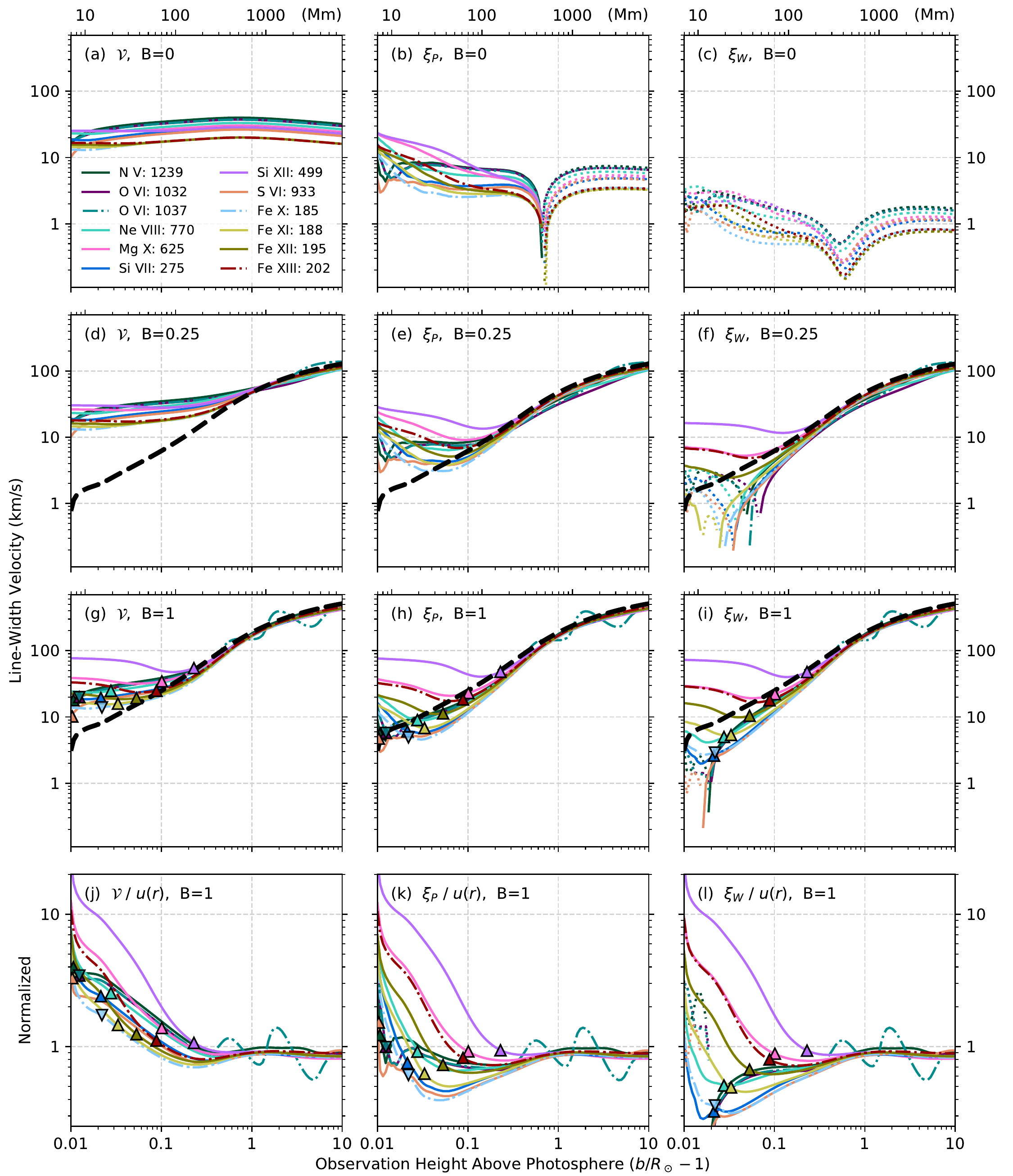}
	\begin{center}
		\caption{Each row presents line-width results from models with increasing wind strengths $B = (0,\ 0.25,\mbox{ and } 1)$. The last row shows $B=1$ again, normalized to radial variation of the solar wind speed $u(r)$. The first column shows $\VV$, second column shows $\xi_P$, and third column shows $\xi_W$. Dotted curves represent negative values. Curve colors are defined in Figure \ref{fig:tempResults}. Dashed black lines show $u(r)$.}
		\label{fig:addingWindVelocity}
	\end{center}
\end{figure*}

%\begin{figure*}[]
%	\begin{center}
%		\includegraphics[width=\linewidthb]{AddingWind_velocity_rel.pdf}
%		\caption{Width measurements at different wind strengths $B = (0,\ 0.25, 1)$. First column shows $\VV$, second column shows $\xi_P$, third column shows $\xi_W$. Dotted curves represent negative values. Curve colors are defined in Figure \ref{fig:tempResults}. Top row is in physical units, next two rows are normalized to radial variation of $u(r)$}
%		\label{fig:addingWindVelocity}
%	\end{center}
%\end{figure*}

\subsection{Results Including Solar Wind} \label{sec:windResults} 

In this section we examine the more self-consistent case in which the solar wind outflow is included in both the ionization calculation and the Doppler broadening. Fast solar wind from the poles is constrained to move along the open magnetic field lines, which expand superradially as discussed in Section \ref{sec:superradialExpansion} and shown in Figure \ref{fig:superRadial}. Several different wind strengths were examined by multiplying $u(r)$ by a constant factor $B$. For each case, the 12 emission lines listed in Table \ref{tab:lineChoices} were synthesized for lines of sight over the pole with a range of impact parameters between $b = 1.01R_\odot$ and $b = 11R_\odot$. We present intensity measurements in Appendix \ref{sec:appendix:intensity}, as well as resonant to collisional intensity fractions. 

Figure \ref{fig:addingWindVelocity} shows the quantities $\VV$, $\xi_P$, and $\xi_W$, determined from the line widths for each ion line as described in Section \ref{sec:understandingLines}. Values of $\VV$ are in the left column, $\xi_P$ is shown in the center, and $\xi_W$ is on the right. Recall that $\xi$ is found by taking the measured width of the spectral line $\VV$ and subtracting a modeled thermal component $V_{th}$. For $\xi_P$ we use the POS value of $T_i$, and for $\xi_W$ we use the weighted temperature $\wt$, as a proxy for an observed line temperature retrieved from some other method. Each row displays increasing wind strengths $B = (0,\ 0.25,\mbox{ and } 1)$, and the last row normalizes the $B=1$ curves by the radial variation in $u(r)$. As before, the peak radii $R_p$ are marked as triangles.
\begin{figure}[]
	\begin{center}
		\includegraphics[width=\linewidthb]{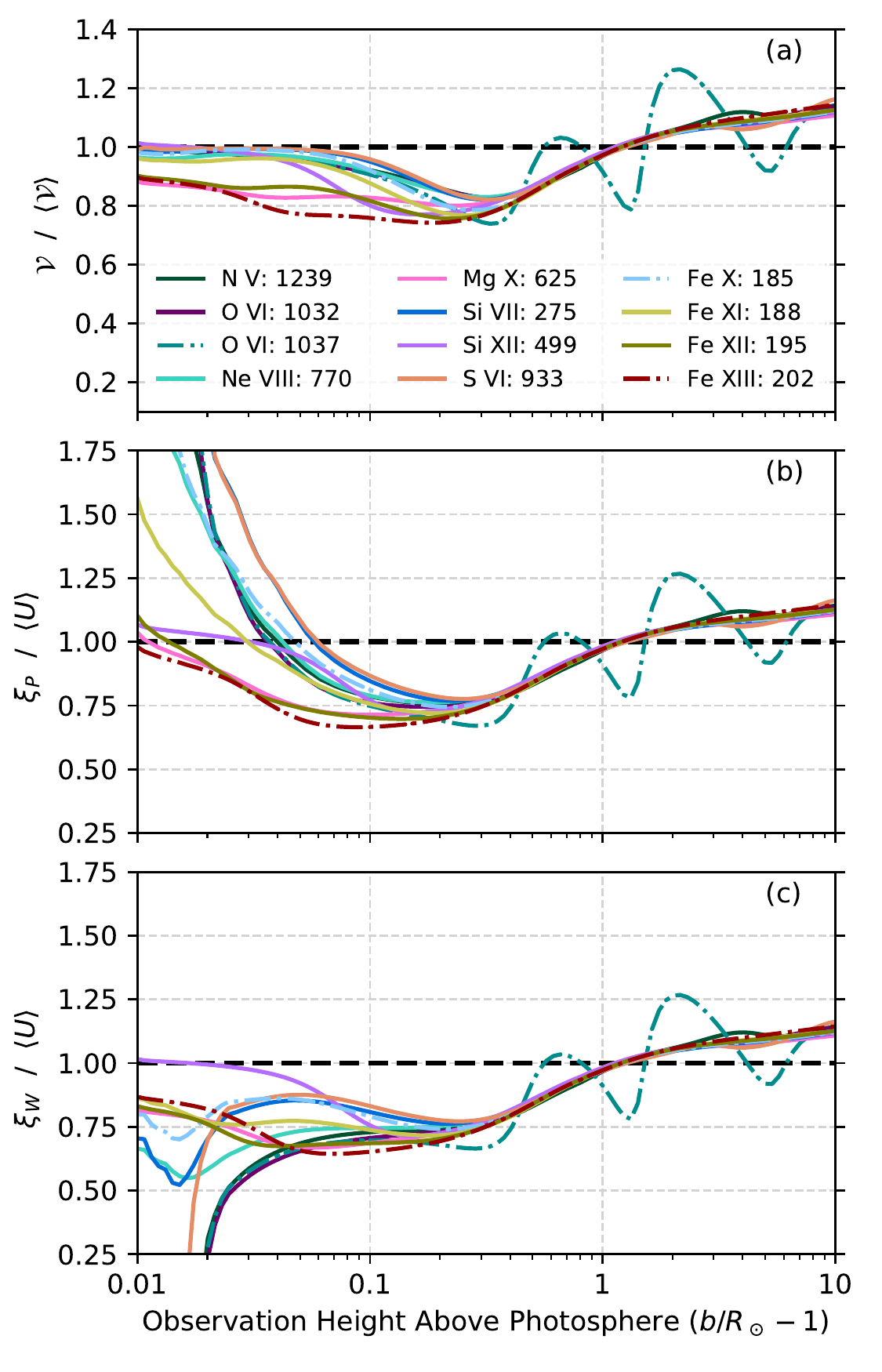}
		\caption{For the $B=1$ case from Figure \ref{fig:addingWindVelocity}(g)-(l): (a) Measured $\VV$ compared to model $\vu$. (b) Measured $\xi_P$ compared to model $\uu$. (c) Measured $\xi_W$ compared to model $\uu$.}
		\label{fig:windRat}
	\end{center}
\end{figure}

In Figure \ref{fig:addingWindVelocity}(a), Figure \ref{fig:addingWindVelocity}(b), and Figure \ref{fig:addingWindVelocity}(c), we examine the same $B=0$ model as discussed in Section \ref{sec:thermalResults} except this time displayed in velocity units. In this flow-free case, one would hope to to be able to recover $\xi=0$, thereby isolating the thermal component $\Tau$. Comparing Figure \ref{fig:addingWindVelocity}(a) and Figure \ref{fig:addingWindVelocity}(b) shows that $\xi_P$ does not do a perfect job of removing the thermal component from the measurement, even in this case with no bulk flows. Above the height of maximum temperature, the POS is the hottest part of the LOS. The cool foreground and background cause a slight reduction in the observed temperature, causing the value of $T_i(r)$ to be an over-correction. Below this height, the hot foreground and background increase the temperature broadening relative to the cool POS, leading to under-correction when constructing $\xi_P$. Figure \ref{fig:addingWindVelocity}(c) shows that $\xi_W$, which uses a temperature inferred from the observation, does a better job of removing the thermal component in this case: The magnitude of $\xi_P$ is on the order of 10 km/s, while $\xi_W$ is closer to 1 km/s. These numbers could be thought of as uncertainties inherent in the measurement; one cannot know $\xi$ to better than these values.

Figure \ref{fig:addingWindVelocity}(a) shows $\VV$ for the flow free case, Figure \ref{fig:addingWindVelocity}(d) shows a reduced wind speed case at $B=0.25$, and Figure \ref{fig:addingWindVelocity}(g) shows the full $B=1$ case. In the upper corona, where all ions have a shared nonthermal velocity that dominates the observation, the measurements $\VV$ tend to be similar in value, and to approximate the POS wind speed $u(r)$ quite well. This is surprising, however, because the solar wind is pointed perpendicular to the LOS in the POS (see Figure \ref{fig:superRadial}), and conventional wisdom would expect very little contribution of the solar wind to the line broadening. In fact, the solar wind has a strong effect on the widths at all heights. Just as in the flow free case, the measured values appear to track the POS value in the upper corona until they plateau. Because the POS is evacuated below each ion's peak radius, observations taken below it are dominated by emissivity at or just below that height. This means that observations taken in the low corona, below where the solar wind has become significant, are still affected by the presence of solar wind in the LOS foreground and background. 

The oscillatory behavior in the \ion{O}{6} line is explained by Doppler pumping. As the velocity of the solar wind increases with height, light from the adjacent \ion{C}{2} 1037 line is scattered by the \ion{O}{6} ions. This causes excess emissivity in the foreground and background of the observation, where the solar wind is much stronger. See Appendix \ref{sec:studyingR} for a full discussion of resonant scattering and Doppler pumping.

In the flow-free case shown in Figure \ref{fig:tempResults}(b), the peak radii (marked by triangles) tend to all occur at the height where the measured temperature has deviated by about 10\% from the POS value, making them a good indicator for where a measurement should track the POS value. In Figure \ref{fig:addingWindVelocity}(j) this does not appear to be the case, with triangles appearing at a range of values of the ratio between 1 and 4. Figure \ref{fig:addingWindVelocity}(k) reveals that the expected behavior is recovered when looking at $\xi_P$, while $\xi_W$ is an over-correction. This makes sense as $\xi$ represents only the non-thermal width, which is most comparable to the POS $u(r)$.

\begin{figure*}[ht!]
	\centering
	\includegraphics[width=\linewidth]{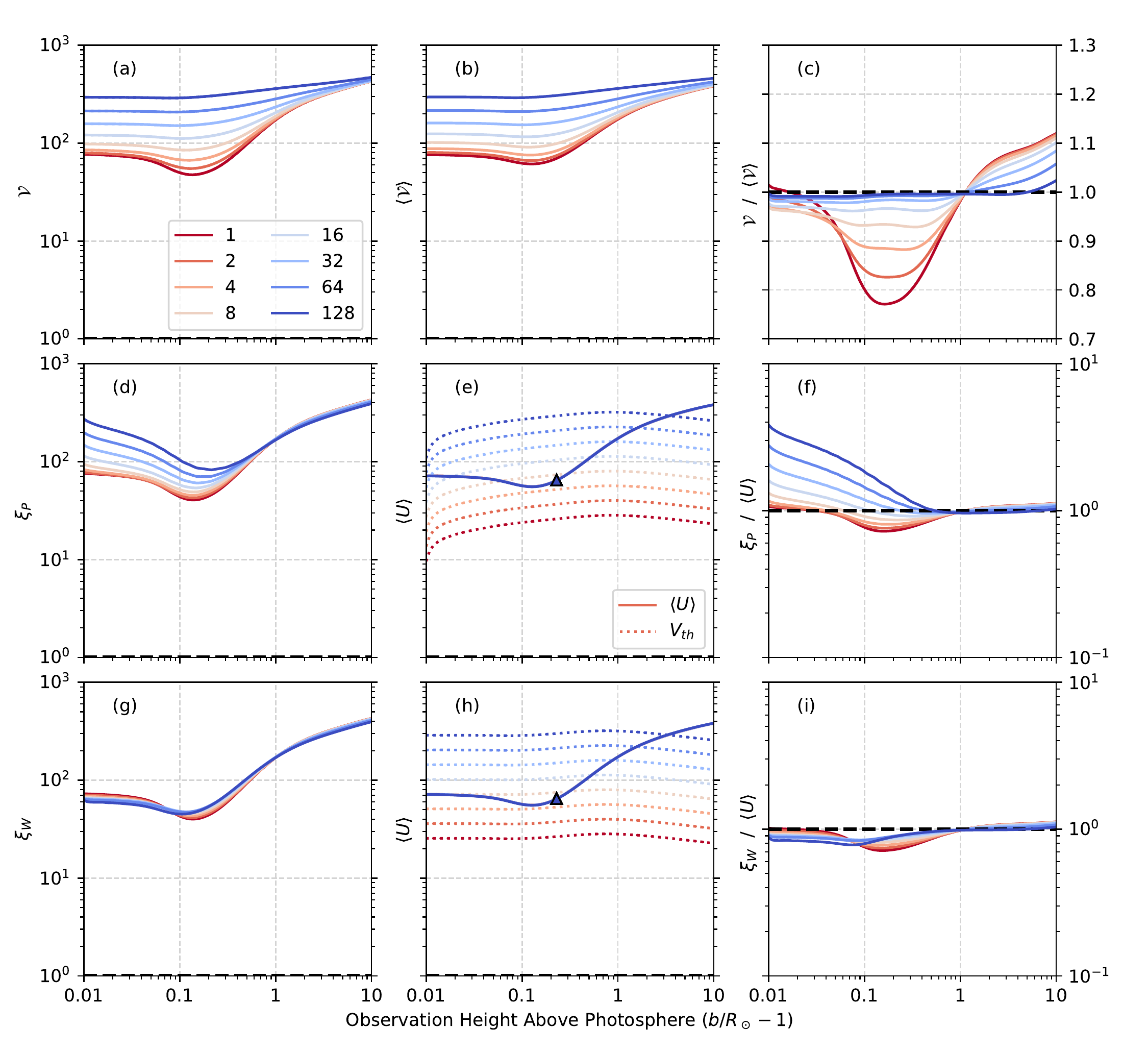}
	\caption{Line width measurements for \ion{Si}{12} 499.406 as a function of preferential ion heating factor $C$. First row shows the full line-width, next two rows examine the two approaches for $\xi$. First column shows reduced line parameters, second column shows LOS modeled parameters, third column shows their ratio. Dotted lines in Panels (e) and (h) show the $V_{th}$ that was subtracted to make their respective $\xi$.}
	\label{fig:boostRat}
\end{figure*}

It is important to point out that making the correction from $\VV$ to $\xi$ only allows the peak radii to be used to determine the domain above which the measurements will match the POS. It does not remove the plateauing effect of the peak radii, as shown by Figure \ref{fig:addingWindVelocity}(k) and Figure \ref{fig:addingWindVelocity}(l). The plateau effect is also highly ion dependent due to the different $R_p$ of each ion: When comparing Figures \ref{fig:addingWindVelocity}(d) and  \ref{fig:addingWindVelocity}(g), the \ion{Si}{12} (violet) line at $b=0.01 R_\odot$ is significantly broadened in the case with stronger wind, but the \ion{S}{6} (orange) line is unchanged.  Clearly, care must be taken to verify that observations are not being interpreted naively below the peak radius of a given ion, where they are not a linear spatial probe of the plasma.

It should be noted that Figures \ref{fig:addingWindVelocity}(g), \ref{fig:addingWindVelocity}(h), and \ref{fig:addingWindVelocity}(i) clearly show a decreasing line-width with height in the low corona. There also seems to be a pronounced dip in the measurements just below the peak radii. The plateau effect could therefore act as a confounding variable when interpreting a decreasing width as an indicator that Alfv\'en waves are being damped in the corona \citep[see, e.g.,][]{Hahn2013}. 

%We use Equation \ref{eq:losAverageV} to construct the quantity $\vu$, which is expected to behave like $\VV$ independent of line width measurements. We follow a similar procedure using Equation \ref{eq:losAverageU} to construct the quantity $\uu$, which should behave like the nonthermal width $\xi$. 

Next we examine the performance of our modeled quantities $\vu$ and $\uu$. Figure \ref{fig:windRat}(a) shows the observed $\VV$ compared to the model $\vu$, and Figures \ref{fig:windRat}(b) and \ref{fig:windRat}(c) show the nonthermal widths $\xi_P$ and $\xi_W$ compared to $\uu$. These models tend to match the simulated observations to within about 25\% above $0.1 R_\odot$, which is quite good, but it should be noted that more work is required to fully understand the width of the spectral lines as a function of LOS properties in this case. 

The low heights, where the thermal component is significant, are different in each panel of Figure \ref{fig:windRat}, while the nonthermally dominated lines in the upper corona are unchanged. It is clear that $\xi_P$ is too broad in the lower corona, indicating that subtracting off the POS temperature is not sufficient to remove the thermal component from $\VV$ (consistent with the plateau effect). However, $\xi_W$ continues to be an over-correction, especially for lines with very low peak radii. While Figure \ref{fig:tempResults}(c) shows that $\wt$ is an excellent model for the thermal effect on the line width in a flow-free case, it does not appear to be a perfect temperature to subtract from $\VV$ in order to recover something like $\uu$. Equation (\ref{eq:losAverageV}) is a good model for $\VV$ only to within 25\%, indicating that there may be additional dependencies required for a more complete version of Equation (\ref{eq:decompose}). The fact that $\VV$ matches $\vu$ better than either of the $\xi$'s can match $\uu$ is interesting, as it could indicate that this method of subtracting off a modeled or measured thermal component isn't feasible in practice. At the very least, it reiterates the lesson that forward-modeled quantities do well at matching observations, while trying to invert an observation can lead to pitfalls. 

\subsection{Results With Preferential Ion Heating} \label{sec:boostResults} 
In this section we analyze the effect of preferential ion heating on observations of spectral widths. There is a lot of work in the literature indicating that it is unlikely that all the ions have the same temperatures due to collisionless kinetic effects \citep[see, e.g.,][]{Hollweg2002, Marsch2006, Cranmer2008a, Chandran2010}.  We examine the effect of preferential ion heating on observed spectral lines by multiplying the baseline ZEPHYR ion temperature as a function of radial distance $T_i(r)$ with a series of constant boost factors $C$ between 1 and 128. The nominal ZEPHYR solar wind speed ($B=1$) was used. For each case, the 12 emission lines listed in Table \ref{tab:lineChoices} were synthesized at 80 impact parameters between $b = 1.01R_\odot$ and $b = 11R_\odot$. Future work will explore temperature anisotropy ($T_{||}\neq T_{\perp}$), but for simplicity we use isotropic temperatures for now.

Figure \ref{fig:boostRat} shows representative results for \ion{Si}{12} 499.406, similar in form to Figure \ref{fig:windRat}. Each curve represents a different boost factor $C$, with hotter temperatures being more blue. Figures \ref{fig:boostRat}(a), \ref{fig:boostRat}(b), and \ref{fig:boostRat}(c) demonstrate that $\VV$ and $\vu$ approximate each other to within twenty percent everywhere, which is true for all ions that we modeled. The dip in the line widths seen in Figure \ref{fig:boostRat} for cooler models is also seen in all ions, with the $C=128$ case tending to have excellent agreement as the lines are so thermally dominated, and the $C=1$ case tending to have a minima around $b=1.3$.

Figures \ref{fig:boostRat}(d) and \ref{fig:boostRat}(g) show the results of subtracting off either the POS temperature or the weighted temperature, as described in Section \ref{sec:understandingLines}. The $V_{th}$ used for each case can be seen in Figures \ref{fig:boostRat} (e) and \ref{fig:boostRat}(h), alongside the emissivity-weighted projected LOS velocity $\uu$. Note that there is no difference in $\uu$ for different values of $C$, as it is a non-thermal quantity. The only exception to this that we observe occur in the Oxygen lines, where the oscillation in the upper parts of the \ion{O}{6} 1037 line appears for the cooler models. Looking at the $V_{th}$ curves, notice that there are plateaus in the weighted temperature case below $R_p$, which are not present in the POS case, but do appear in Figures \ref{fig:boostRat}(a) and \ref{fig:boostRat}(b). For most ions, $\xi_P$ tends to under-correct and show residual thermal dispersion, while $\xi_W$ does a better job of isolating the non-thermal component (or slightly over-correct), as seen in Figures \ref{fig:boostRat} (f) and \ref{fig:boostRat}(i).

\begin{figure}
	\centering
	\includegraphics[width=\linewidth]{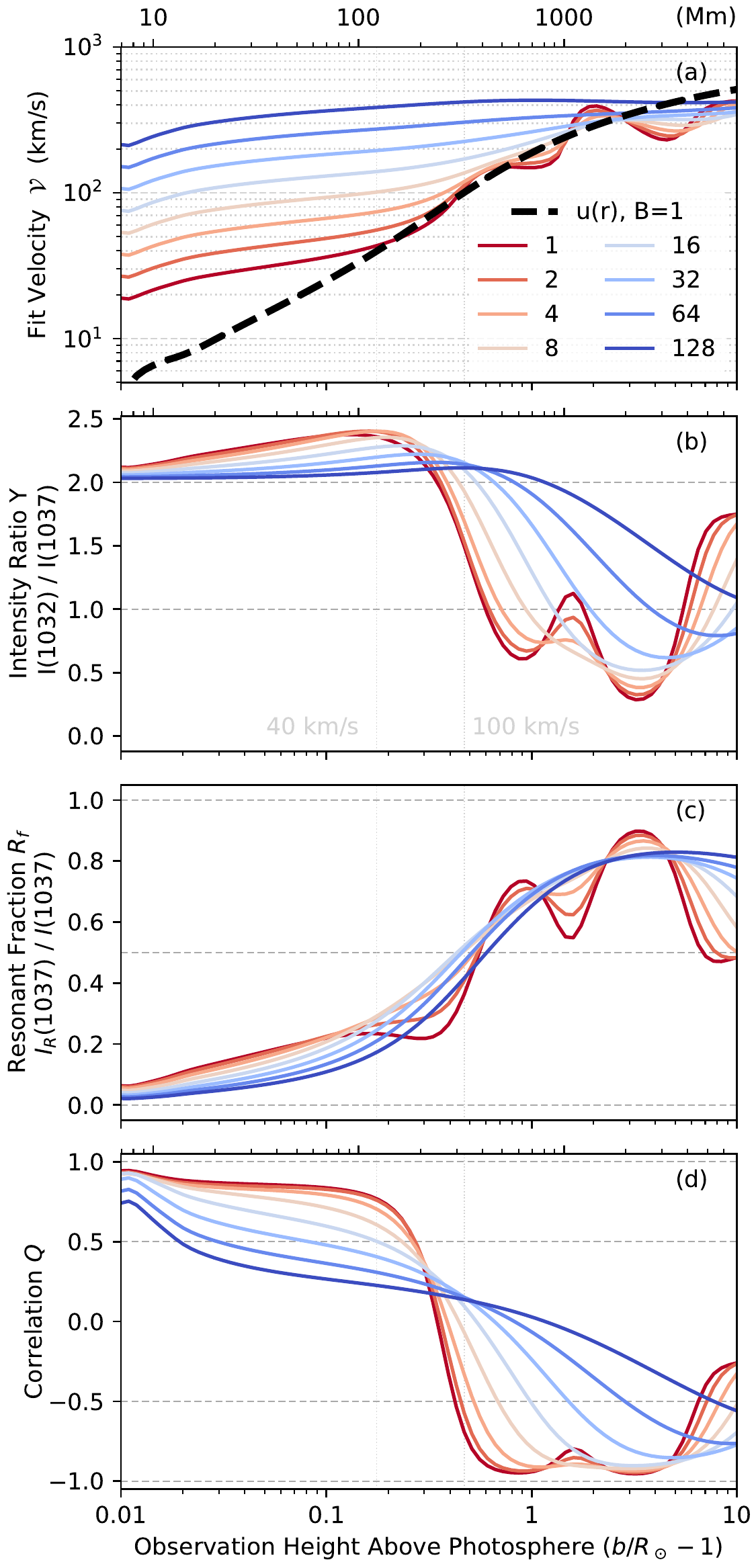}
	\caption{The effect of preferential ion heating on the spectral line O VI 1037. Colors correspond to boost factor $C$. Vertical grey lines mark height where $u(r) = 40$ km/s and $100$ km/s. (a) Line fit velocity $\VV$. (b) Intensity ratio $\Upsilon = I(1032)/I(1037)$. (c) Resonant fraction $R_f$. (d) Correlation $Q= \Upsilon/R_f$.} %[Should Overplot Real Data]
	\label{fig:boostPlot}
\end{figure}

In Figure \ref{fig:boostPlot}, we proceed to examine the \ion{O}{6} 1032 and 1037 lines more closely.   Figure \ref{fig:boostPlot}(a) shows the measured line width $\VV$ for \ion{O}{6} 1037 . As the temperature gets higher, the thermal velocity component increases proportionately. The solar wind velocity component is present at most temperatures, though the highest temperature models are completely dominated by the thermal component. This is true for measurements of all ions, not just oxygen. 

We include two strong pumping lines in the incident spectrum for \ion{O}{6} 1037, which leads to oscillations in the measurement (see Appendix \ref{sec:studyingR}). As the temperature is raised, these structures in the observations are smoothed out due to the heavily broadened lines.  In this case, even a relatively moderate amount of preferential heating $(C = 4 - 8)$ has completely removed the behavior. Note that the double peak is not seen in the literature \citep{Cranmer2008a, Antonucci2012}. This could indicate that preferential heating is in fact occurring, as it does not appear in our simulations with $C$ greater than about 6. 

Figure \ref{fig:boostPlot}(b) shows the ratio of the integrated line intensities $\Upsilon=I$(1032)/$I$(1037), and Figure \ref{fig:boostPlot}(c) shows the resonant fraction of the spectral line $R_f=I_R(1037)/I(1037)$. In the lower regions of the corona, these quantities are thought to be correlated, with a value of the line ratio close to 2 indicating a collisionally dominated line, and a value closer to 4 indicating a scattering dominated line \citep{Kohl1982}. A formal comparison is plotted in Figure \ref{fig:boostPlot} (d), which shows 
\begin{equation}
Q = \frac{1}{2}\left(\frac{I(1032)}{I(1037)} - 2\right) \frac{I(1037)}{I_R(1037)},
\end{equation}
which simply rescales $\Upsilon$ and then takes the ratio of that with the resonant fraction. This function was chosen to illustrate how well correlated the behaviors of these two quantities are with one another. We find a strong correlation of these behaviors for the cooler models, with hotter models correlating less. In the upper corona $Q$ is close to -1, showing a much more precise anti-correlation than we expected. This because the increasing $I_R(1037)$ due to the Doppler pumping appears inversely in both quantities, in the denominator of $\Upsilon$ and the numerator of $R_f$. In other words, the increasing fraction of resonant light $R_f$ increases the total intensity of that line $I(1037)$, decreasing the line ratio $\Upsilon$. 

\cite{Noci1987} pointed out that Doppler pumping will cause the \ion{O}{6} 1037.6 line to overlap with the \ion{C}{2} 1037.0 line at a relative velocity of about 100 km s$^{-1}$, which implies that when the solar wind has accelerated to 100 km s$^{-1}$, the Oxygen line ratio should drop below 2 as 1037 brightens significantly. Our simulation is roughly consistent with that behavior. In Figure \ref{fig:boostPlot}(b), the height (and wind velocity) where $\Upsilon$ crosses two is a fairly strong function of temperature, with a velocity of $100$ km/s best matching a model with a boost factor $C$ of $\approx 12$. It seems that the correlation $Q$ begins to drop at a height where the wind speed is around $40$ km/s ($b\approx1.2$) for the coolest models.

%\clearpage
%%%%%%%%%%%%%%%%%%%%%%%%%%%%%%%%%%%%%%%%%%%%%%%%%%%%%%%%%%%%%%%%%%%%%%%%%%%%%%%%%%%%%%%%%%%%%%%%%%%%%%%%%%%%%%%%%%%%%%%%%%%%%%%%%%%%%%%%%%%%%%%%%%%%%%%%%%%%%%%%%%%%%%%%%%%%%%%
%%%%%%%%%%%%%%%%%%%%%%%%%%%%%%%%%%%%%%%%%%%%%%%%%%%%%%%%%%%%%%%%          End of Paper Sections           %%%%%%%%%%%%%%%%%%%%%%%%%%%%%%%%%%%%%%%%%%%%%%%%%%%%%%%%%%%%%%%%%%%%%
%%%%%%%%%%%%%%%%%%%%%%%%%%%%%%%%%%%%%%%%%%%%%%%%%%%%%%%%%%%%%%%%%%%%%%%%%%%%%%%%%%%%%%%%%%%%%%%%%%%%%%%%%%%%%%%%%%%%%%%%%%%%%%%%%%%%%%%%%%%%%%%%%%%%%%%%%%%%%%%%%%%%%%%%%%%%%%%
\section{Discussion/ Summary} \label{sec:discussion}

In this work we have examined the relationship between quantities observed through long lines of sight and the true radial variation of the plasma near the Sun. We have discovered that Line of Sight effects can be quite large, leading to both systematic errors in measurement and an observed functional form that is completely different than the true variation of the quantity. When attempting to spectroscopically measure plasma parameters as a function of height above the solar surface, we suggest one use spectral lines from ions that are increasing in density with depth (in other words, only observe above $R_p$ for a given ion). Below $R_p$, where a given ion's density is maximized, the observation will plateau and no longer match the radial values in the POS. We refer to this as the plateau effect. 

In our simulated observations, we find that the solar wind has dominant effect on spectral widths, even at heights far below where the solar wind is thought to be significant (due to the plateau effect). Therefore, when interpreting the widths of spectral lines from the optically-thin corona, care should be taken to consider the solar wind as a source of broadening. This may reduce the amount of preferential ion heating required to match some observations in future models. We determined that targeting the POS value of $T_i$ to create $\xi_P$ tends to under-correct the measurement. On the other hand, using a LOS temperature such as $\wt$ does a better job at removing the thermal component of the line. 

We determined that it is easy to overly truncate these types of simulations. In general, several solar radii in and out of the POS should be considered, or the LOS effects may be under-simulated by truncating portions of the LOS that contribute to the observation. In resonant scattering calculations, it is also important to include a continuum component in the incident light profile, or the effects of Doppler dimming can be over-simulated.  See Appendix \ref{sec:studyingR} for an in-depth analysis. 

%Appendix \ref{sec:multiIon} discusses our rejection of a method for using multi-ion information to determine the thermal component of a spectral line. See Appendix \ref{sec:appendix:intensity} for a discussion of how the strength of the solar wind was found to strongly modulate the intensity of the resonantly scattered component of many spectral lines, alongside simulated intensity results.

Future work with GHOSTS will involve adding waves and inhomogeneities to the model and coming up with new analysis tools to interpret their effect on the simulated observations. We would also like to include effects such as photoionization, activity-cycle variations in the solar-disk spectrum, and non-Maxwellian velocity distributions. Additional work on calculating the populations of excited (non-resonant) energy levels, as well as the strength of forbidden transitions, would allow us to simulate DKIST lines. Work should be done in the future to see if any of these	results apply to temperature measurements that are derived from other methods than spectral line widths, such as off-limb rotational tomography involving EUV imaging \citep[e.g.][]{Frazin2005, Nuevo2015, Lloveras2020}.

%%%%   Footer   %%%%%%%%%%%%%%%%%%%%%%%%%%%%%%%%%%%%%%%%%%%%%%%%%%%%%%%%%%%%%%%%%%%%%%%%%%%%%%%%%%%%%%%%%%%%%%%%%%%%%%%%%%%%

\acknowledgments
{The authors gratefully acknowledge Sam Van Kooten, Ben Boe, and Mike Hahn for many valuable discussions, as well as countless others at poster sessions over the past few years. We are also very grateful to the anonymous referee, whose review has allowed us to clarify many parts of this paper. This work was supported by the National Aeronautics and Space Administration (NASA) under grants NNX15AW33G and NNX16AG87G, and by the National Science Foundation (NSF) under grants 1540094 and 1613207.  Thanks also to Juri Toomre's Laboratory for Computational Dynamics for providing us with computing time. }
\software
{NASA's Astrophysics Data System (ADS), 
	Python version 3.6.5 \citep{Millman2011}, 
	numpy \citep{VanderWalt2011}, 
	scipy \citep{Oliphant2007}, 
	scikit-image \citep{VanderWalt2014}, 
	MPI4py \citep{Dalcin2011},
	matplotlib \citep{Hunter2007}, and
	CHIANTI v8 \citep{DelZanna2015}}

%\facilities{facility ID, facility ID, facility ID, ... }

%%%%   Appendix   %%%%%%%%%%%%%%%%%%%%%%%%%%%%%%%%%%%%%%%%%%%%%%%%%%%%%%%%%%%%%%%%%%%%%%%%%%%%%%%%%%%%%%%%%%%%%%%%%%%%%%%%%%

\clearpage
 \appendix 
\section{Impact of the Choice of Incident Spectrum on Resonant Broadening} \label{sec:studyingR}
The solar wind broadens coronal spectral lines due to large-scale LOS-projected velocity components. For the resonantly scattered component of the spectral line, this broadening is modulated by Doppler dimming and pumping, in which a relative velocity between the emitting and scattering particles decreases or increases the scattering cross section, respectively. This changes the significance of the foreground and background of the line of sight as a function of solar wind velocity, which will alter the value of LOS measurements \citep{Withbroe1982, Noci1987, Kohl2006}. One example of this is the oscillations in the \ion{O}{6} 1037 line width in Figure \ref{fig:addingWindVelocity} and Figure \ref{fig:windRat}: when the foreground and background are more dominant, the solar wind at that location contributes more broadening to the total line. 

\begin{figure}[b]
	\centering
	\includegraphics[width=0.5\linewidth]{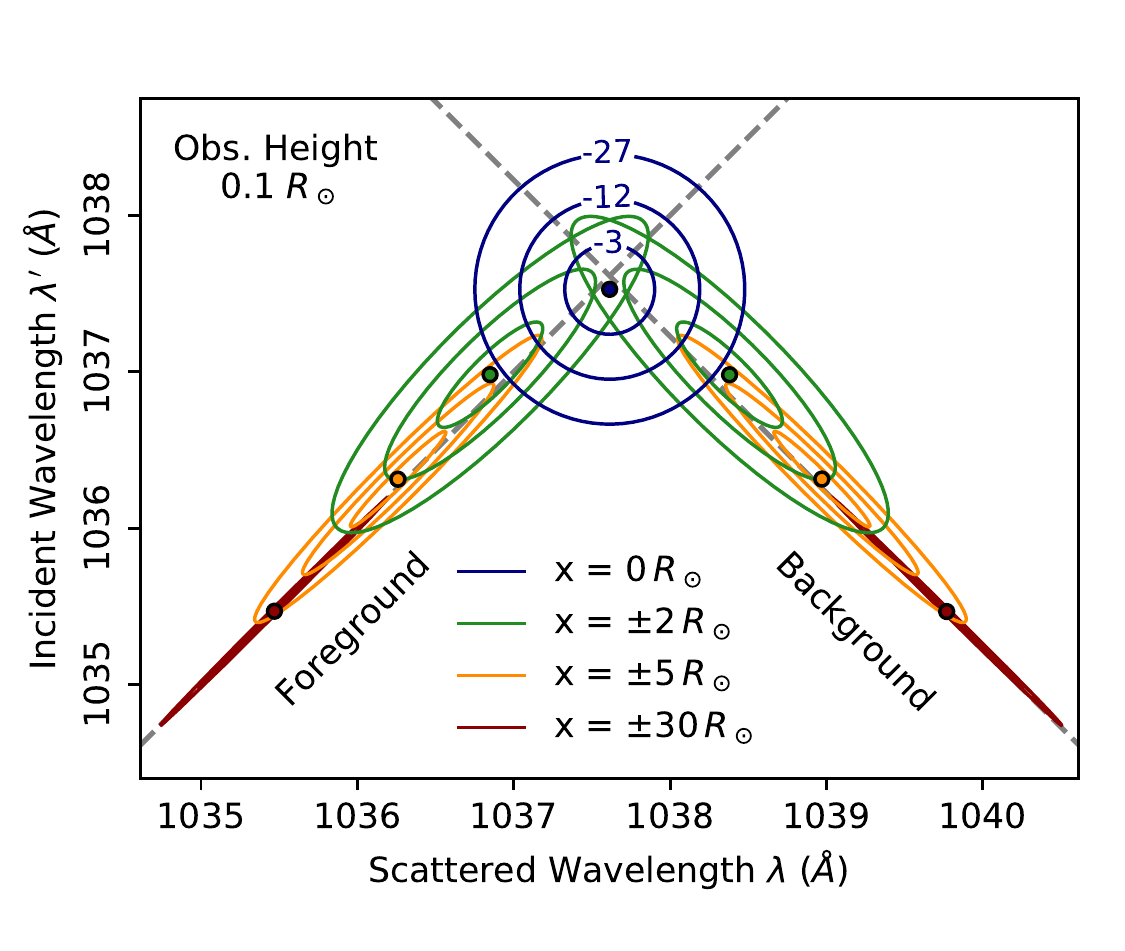}
	\caption{Contours of the redistribution function $\log_{10}({\cal R})$ for \ion{O}{6} 1037 at several points along a LOS simulated at $b=1.1R_\odot$.}
	\label{fig:Rfunc}
\end{figure}

To accurately compute the effects of Doppler dimming and pumping on the measured line widths, it is important to specify the incident spectrum carefully. Figure \ref{fig:Rfunc} shows the redistribution function ${\cal R}$, defined in Equation (\ref{eq:resonantEmiss}). At each point along the LOS, the wavelength dependencies of ${\cal R}$ are quasi-Gaussians in the two-dimensional $\{ \lambda, \lambda' \}$ plane. The abscissa and ordinate of this plot are the scattered and incident wavelengths, which we can write using dimensionless frequency shifts (as in Equations \ref{eq:rezemissivity} and \ref{eq:dimensionlessShift})) as
\begin{equation}
x \, = \, \frac{\nu - \nu_0}{\Delta \nu} = \frac{\lambda-\lambda_0}{\Delta\lambda}
\,\,\,\,\,\,\,\,\,
\mbox{and}
\,\,\,\,\,\,\,\,\,
y \, = \, \frac{\nu' - \nu_0}{\Delta \nu} = \frac{\lambda'-\lambda_0}{\Delta\lambda} \,\, ,
\end{equation}
where 
\begin{equation}
\frac{\Delta\lambda}{\lambda} \approx \frac{\Delta\nu}{\nu}
\end{equation}

Figure \ref{fig:Rfunc} illustrates the LOS dependence of ${\cal R}$ for a representative LOS with $b = 1.1 \, R_{\odot}$. Near the POS (i.e., $x \approx 0$) the scattering angle $\theta$ is close to $90^{\circ}$ and the LOS-projected wind speed ${\bf u} \cdot \hat{\bf n}$ is close to zero. Thus, the quasi-Gaussian part of the redistribution function is given approximately by
\begin{equation}
{\cal R} \, \propto \, \exp \left[ - \left( y - \frac{u}{v_{th}}
\right)^2 - x^2 \right]
\end{equation}
which is shown as concentric circular contours, shifted slightly down from the intersection of the two diagonal lines described below.

On the other hand, in the extreme foreground ($\theta \rightarrow 0^{\circ}$) and background ($\theta \rightarrow 180^{\circ}$) along the LOS, the quantity $\beta = \sin\theta$ is close to zero, and the term containing $\beta$ in the denominator dominates the exponential. In those limits, the $u$-dependent Doppler dimming terms cancel out and
\begin{equation}
{\cal R} \, \propto \, \exp \left[ - \left( \frac{x \pm y}{\beta}
\right)^2 \right]
\end{equation}
where the upper sign corresponds to the background and the lower sign corresponds to the foreground. Because $\beta \approx 0$, the only non-vanishing parts of ${\cal R}$ correspond to $x \pm y \approx 0$, which are shown in Figure \ref{fig:Rfunc} by the two diagonal dashed lines. Computationally, this can lead to distortions in the results if the resolutions of the $\lambda$ and $\lambda'$ axes are not sufficiently high, as the matrix becomes a sparse, broken diagonal. We find that spectral resolution of $N_{\lambda}= 200$ and $N_{\lambda'}=250$ for the incident and scattered wavelength axes is sufficient for a LOS with $b=11R_\odot$, and which extends to $x=\pm75R_\odot$, (defined in Section \ref{sec:spectralLines} as $S(11) = 75$).

\begin{figure*}[]
	\centering
	\includegraphics[width=0.87\textwidth]{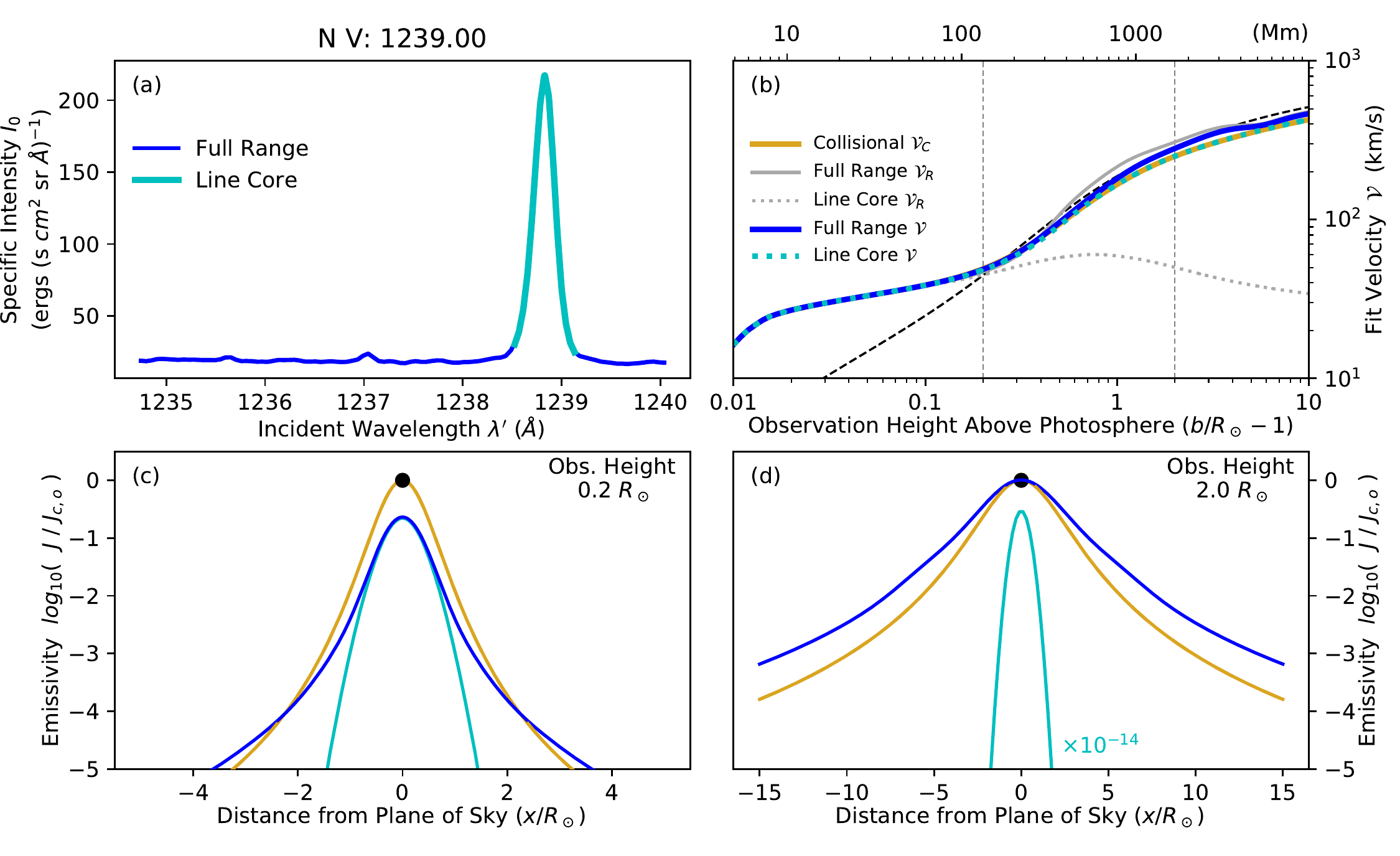}
	\caption{(a) Two choices for the incident spectrum $\lambda'$ used in redistribution for \ion{N}{5} 1239. (b) The effect of this choice on $\VV$, $\VV_R$, and $\VV_C$. The black dashed line is the wind speed $u(r)$. (c) and (d) Relative emissivity $J(x)$ for LOS at observation heights marked by vertical gray lines in (b). Curves are normalized to $J_C(0)$ (the black point), and the cyan curve is multiplied by the indicated factor.}
	\label{fig:incident2}
\end{figure*}

\begin{figure*}[]
	\centering
	\includegraphics[width=0.87\textwidth]{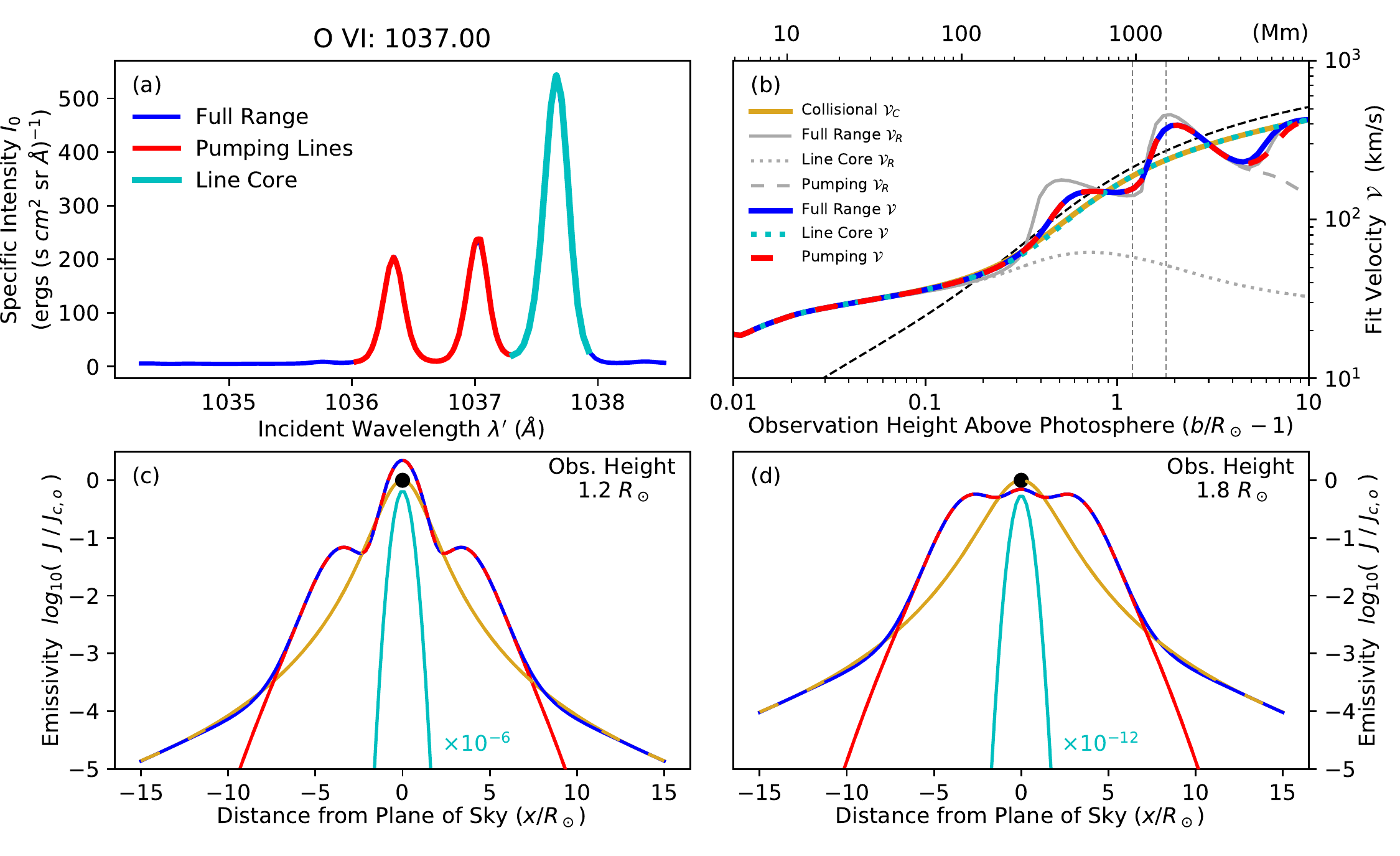}
	\caption{(a) Three choices for the incident spectrum $\lambda'$ used in redistribution for \ion{O}{6} 1037. (b) The effect of this choice on $\VV$, $\VV_R$, and $\VV_C$. The black dashed line is the wind speed $u(r)$. (c) and (d) Relative emissivity $J(x)$ for LOS at observation heights marked by vertical gray lines in (b). Curves are normalized to $J_C(0)$ (the black point), and the cyan curve is multiplied by the indicated factor.}
	\label{fig:incident3}
\end{figure*}

The behavior of the redistribution function in the two-dimensional $\{ \lambda, \lambda' \}$ plane is important to understand because the numerical integration over $\lambda'$ must be done over the range of frequencies that correspond to the non-vanishing parts of ${\cal R}$. If, instead, the range of numerical $\lambda'$ values was chosen to focus only on the peak of the corresponding emission line in the solar spectrum, it is possible that ${\cal R}$ could be truncated artificially. In other words, it is important to include a continuum around the spectral line, rather than modeling it only as a Gaussian. 

Figures \ref{fig:incident2} and \ref{fig:incident3} show what happens to the observations when different choices are made for the limits of the incident spectrum $I_0$ in Equation \ref{eq:rezemissivity}. Let us first examine Figure \ref{fig:incident2}, which shows results for \ion{N}{5} 1239. Figure \ref{fig:incident2}(a) shows two choices for incident spectra: In the line-core only case (cyan), we simply include the Gaussian component of the photospheric line. In the full-range case (blue), we set the limits such that all parts of the redistribution function ${\cal R}$ greater than $10^{-30}$ are included along the line of sight. Figure \ref{fig:incident2}(b) shows line-widths $\VV$ as a function of observation height. The gold curve shows the width of the collisional component of the line, which is the same for all cases, and the remaining colored curves represent measurements taken on the entire summed spectral line. The two grey curves represent the width of only the resonant component of the line for each case. In the line-core only case, the resonant component of the line is far too narrow, and the total measurement is entirely dominated by the collisional component of the line. When the full spectral range is considered, the true resonant component of the line is revealed to be broader even than the collisional component, and the total line width is affected. 

Figures \ref{fig:incident2}(c) and \ref{fig:incident2}(d) show the LOS dependence of the local emissivities $J_{r}(x)=\int j_r(x,\nu)d\nu$ for the two cases, as well as $J_c$ (in gold), at the heights marked by vertical lines in \ref{fig:incident2}(b). Simply using the line-core causes the resonant emissivity to be localized only to the plane of the sky (the cyan curves), whereas in the full range case the foreground and the background emissivity is much more significant. When there is no continuum, the Doppler-shifting due to the solar wind causes the scattered light profile to no longer overlap with the incident light profile. This leads to no emissivity at all in the foreground and background, as well as a reduction in the POS intensity, which completely changes the LOS emissivity profile. At low heights, the effect on the measurements is subtle, but it gets much stronger in the upper corona where the solar wind speed is significant. This study indicates that it is important to include the full range across which ${\cal R}$ is significant (i.e. including a continuum around the spectral line), or line width measurements will not be accurate. 

Figure \ref{fig:incident3} examines the more complex case of \ion{O}{6} 1037, and we include an additional case (in red) to explore the effect of the two adjacent ``pumping lines." The oscillatory nature of the line widths in Figure \ref{fig:incident3}(b) is caused by Doppler pumping from these two lines, which modify the relative importance of the foreground and background of the LOS. Figures \ref{fig:incident3}(c) and \ref{fig:incident3}(d) again show the LOS dependence of the local emissivities $J(x)$. We believe that the fact that the resonant and collisional emissivities have very similar values in Figure \ref{fig:incident2}(c) is a coincidence. At heights where the spectral line is narrow, the emissivity is highly peaked in the POS, but where it is broad, a large region of the LOS near the POS has around the same intensity. Because the solar wind in the foreground and background has a greater velocity component that is directed into the LOS, the line is broadened.  Simply using the line-core removes this behavior entirely, but notice that for higher measurements, just including the pumping lines is insufficient, and additional continuum is still required to produce accurate results. Above a certain height, the resonant emissivity drops off as $r^{-4}$, just like the collisional emissivity. This is because the collisional emissivity drops as $\rho^2\propto r^{-4}$, and the resonant emissivity drops off as $\rho W(r)\propto r^{-4}$ as long as there is a flat incident continuum wherever ${\cal R}$ is significant.

\begin{figure}[]
	\centering
	\includegraphics[width=0.5\linewidth]{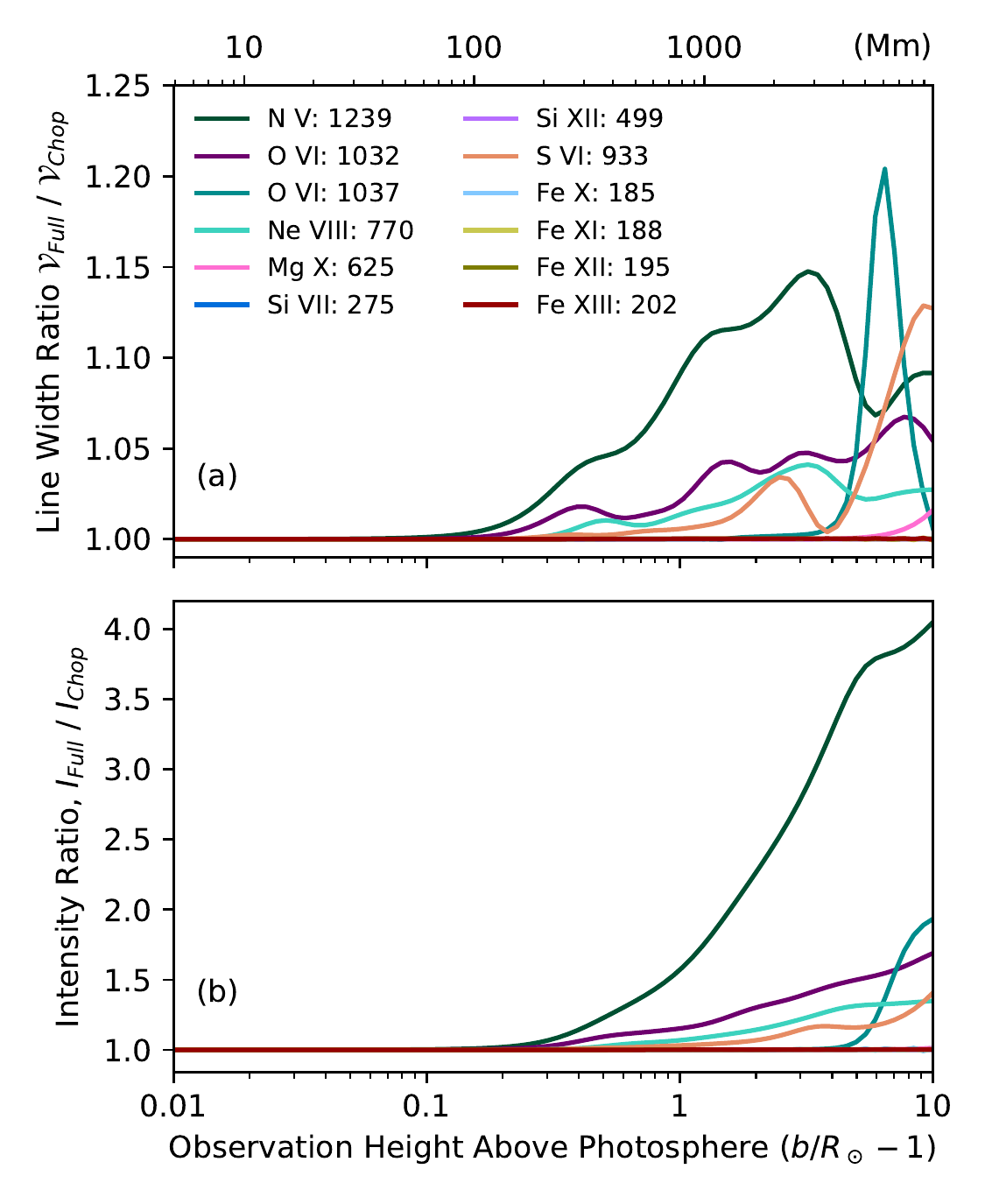}
	\caption{(a) Excess width caused by including continuum. (b) Excess intensity caused by including continuum.}
	\label{fig:incident4}
\end{figure}

%It is clear that $\VV_C$ follows the POS value of the solar wind closely, while the resonant widths $\VV_R$ depend strongly on the choice of limits for $\lambda'$.
Figure \ref{fig:incident4} illustrates the effect of truncating the incident spectrum for each ion we considered. Figure \ref{fig:incident4}(a) shows that the difference in the measured spectral width is on the order of 5-20\%, and Figure \ref{fig:incident4}(b) indicates that the intensity can be affected by up to a factor of four. While this applies to the total spectral line, the resonant component itself can be affected by much larger factors. These effects are far less significant in the low corona than the higher regions, but as technology continues to improve our field of view, it must be considered.

\clearpage
\section{Using Multi-Ion Observations to Understand Spectral Width} \label{sec:multiIon}

As described in Section \ref{sec:understandingLines}, when interpreting the width of a spectral line, it is difficult to disambiguate the thermal width $v_{th}$ and the non-thermal width $\xi$, as the quantities are convolved together into a single width measurement $\VV$. One attempt to do so has been to recognize that the thermal width depends on the mass of the emitting ion, but most models for the non-thermal width do not. By utilizing multi-ion measurements, and assuming that all ions have a common temperature $T_i$ and bulk motion $u$, one can attempt to take advantage of this mass dependence to tease out the temperature component \citep{Seely1997, Moran2003}.  For a set of ions $i$, Equation (\ref{eq:decompose}) can be treated as a linear least-squares fitting model, performing fits to these measured (i.e., simulated) sets of $\VV_i^2$ values as a function of the inverse ion mass. The slope of the fit line is proportional to $\Tau$, and the intercept gives the non-thermal velocity $\xi$. Because the fits are performed in $V^2$, there may occur negative values of the intercept $\xi^2$, which would be unphysical. As a note, it does not seem that the assumption of a common temperature and bulk motion are consistent with observations to date \citep{Hollweg2002, Marsch2006, Cranmer2008a, Chandran2010}. However, these assumptions do hold within the GHOSTS simulation, so we can test the procedure in an idealized environment.

\begin{figure}[b]
	\centering
	\includegraphics[width=\linewidth]{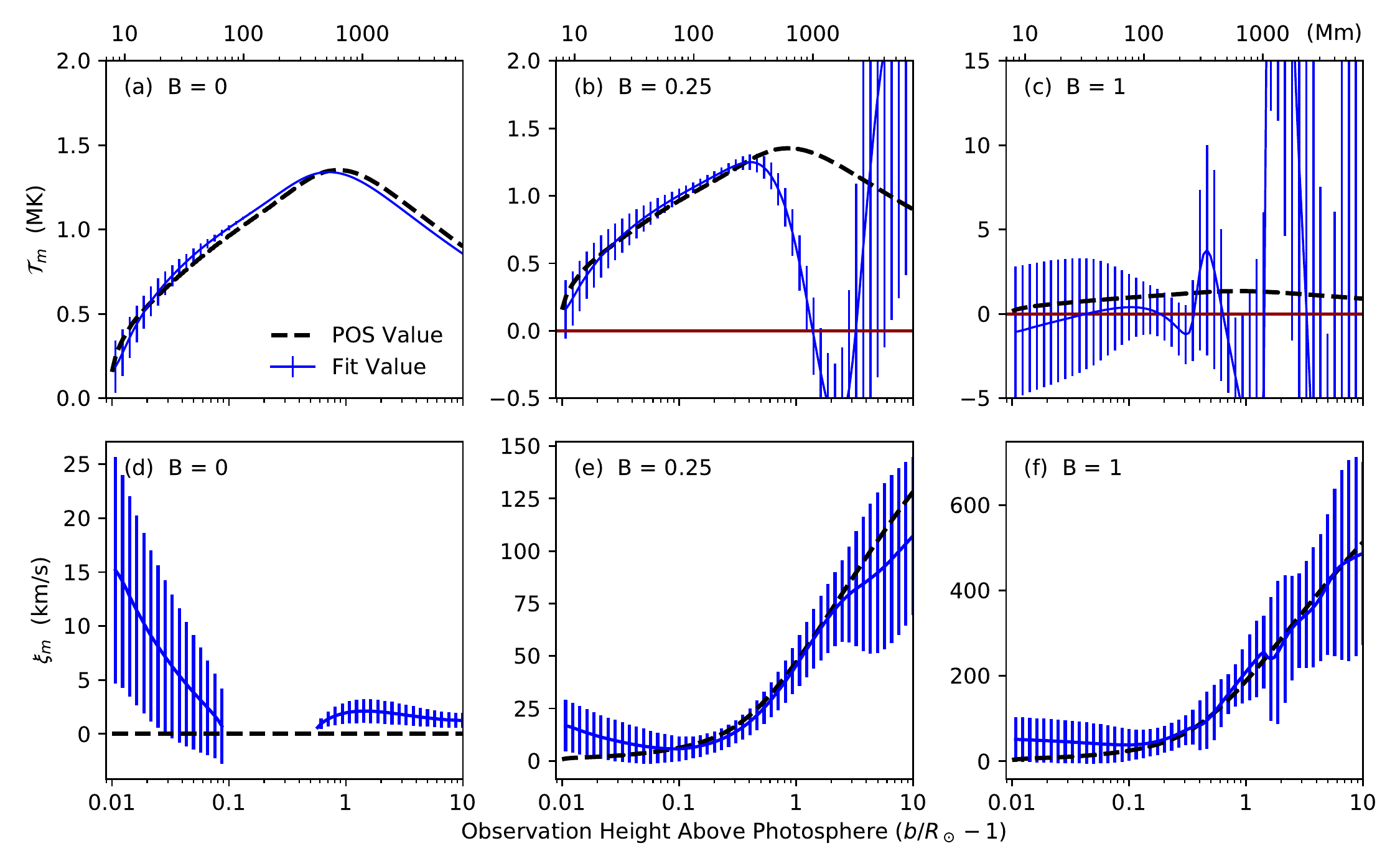}
	\caption{Top row shows common fit-temperature $\Tau_m$, bottom row shows common non-thermal velocity $\VV_m$, each with $1\sigma$ fitting uncertainties. Dashed black lines show POS values of the $T_i$ and $u(r)$, respectively. Each column represents a different wind strength $B$.}
	\label{fig:moranPlot}
\end{figure}

The common fit-temperature $\Tau_m$ and non-thermal velocity $\xi_m$ were computed as a function of observation height for simulations with several values of the solar wind factor $B$. The results are plotted in Figure \ref{fig:moranPlot}. Figure \ref{fig:moranPlot}(a) shows that this procedure performs remarkably well at retrieving $T_i$ for the case with no wind. Across the entire domain, the functional form of the curve is retrieved. In the upper regions, the fit uncertainties are very small because all the ions have common values of $\Tau$. In the lower regions, where the $\Tau$ values start to diverge because of the measurement floors, the precision of the technique is reduced slightly. Nevertheless, the measurement floors have been effectively removed from the observation, with the POS value of $T_i(r)$ well within the error bars. On the other hand, although there should be no nonthermal velocity detected in these observations, Figure \ref{fig:moranPlot} shows that the computed value of $\VV_m$ is on the order of 10 km/s in the lower corona (with large error bars) and 2 km/s in the upper corona, which suggests that this method may not be a reliable way to discriminate between $\Tau$ and $\xi$. The extra non-thermal velocity is likely caused by the temperature plateaus, which violate the assumptions of the fitting model: Even though the ions all share a common $T_i$, their effective $\Tau$ are not the same. The effects of preferential ion heating are sure to exacerbate this problem.

For the cases with solar wind, it seems that even in a noise-free, high dynamic range simulation, the mass-dependent signal that this method attempts to tease out is too small to be measured. Whenever wind was introduced, the value of $\Tau_m$ was unable to approximate the POS value of the wind in the upper corona. On the other hand, $\xi_m(b)$ does a fairly good job of approximating $u(r)$, even when $\Tau_m$ is very wrong. This may not actually be very useful, however, as this method works by fitting a straight line through a set of points, with $\xi_m$ determined by its y-intercept value. Because the mass-dependent thermal width $V_{th}$ is small compared to the wind broadening, this fit line is in essence a horizontal line at the average value of the points. Because this method fails to find $\Tau_m$ in the presence of even relatively modest $(B=0.25)$ solar wind, this method does not provide useful extra information about the observations over just taking the average of the velocity measurements for each ion. Additionally, because it finds spurious non-thermal velocity in the flow free case, which persists in the cases with wind, it can not be trusted to provide $\xi_m$ either. We believe that this type of analysis is unable to be performed reliably in principle.

 %One can attempt to use $\beta_T$ to correct $\Tau_m$ for LOS effects. The red and cyan curves in Figure \ref{fig:correctThermal}(a) were calculated by dividing $\Tau_m$ by $\beta_{T,c}$ or $\beta_{T,r}$, respectively. The correction is only on the order of a few percent, but it is clear that the corrected curves match $T_i$ better in the upper regions of the plot. The lower region isn't corrected as well, because the true weighting functions are different for each ion in that region. In fact, both curves overcorrect the measurements at that height. In this $B=0$ case, it seems that the LOS corrections are most applicable in the upper regions of the corona, where all the ions are behaving similarly.

\section{Intensity Analysis} \label{sec:appendix:intensity}

Absolute intensities are a key measurable quantity that we have not studied very much in this work. Yet as they are generated in the course of the simulation, we are able to provide them here. Because of the semi-empirical nature of our code, we are also able to separate the sources of intensity, and provide a measure of the resonant to collisional fraction for each line as a function of height. This provides an interesting look at the direct effects of Doppler dimming, and may provide the reader with the ability to make more informed choices about which lines might be useful as diagnostics of different processes. 

\begin{figure}[ht!]
	\centering
	\includegraphics[width=\textwidth]{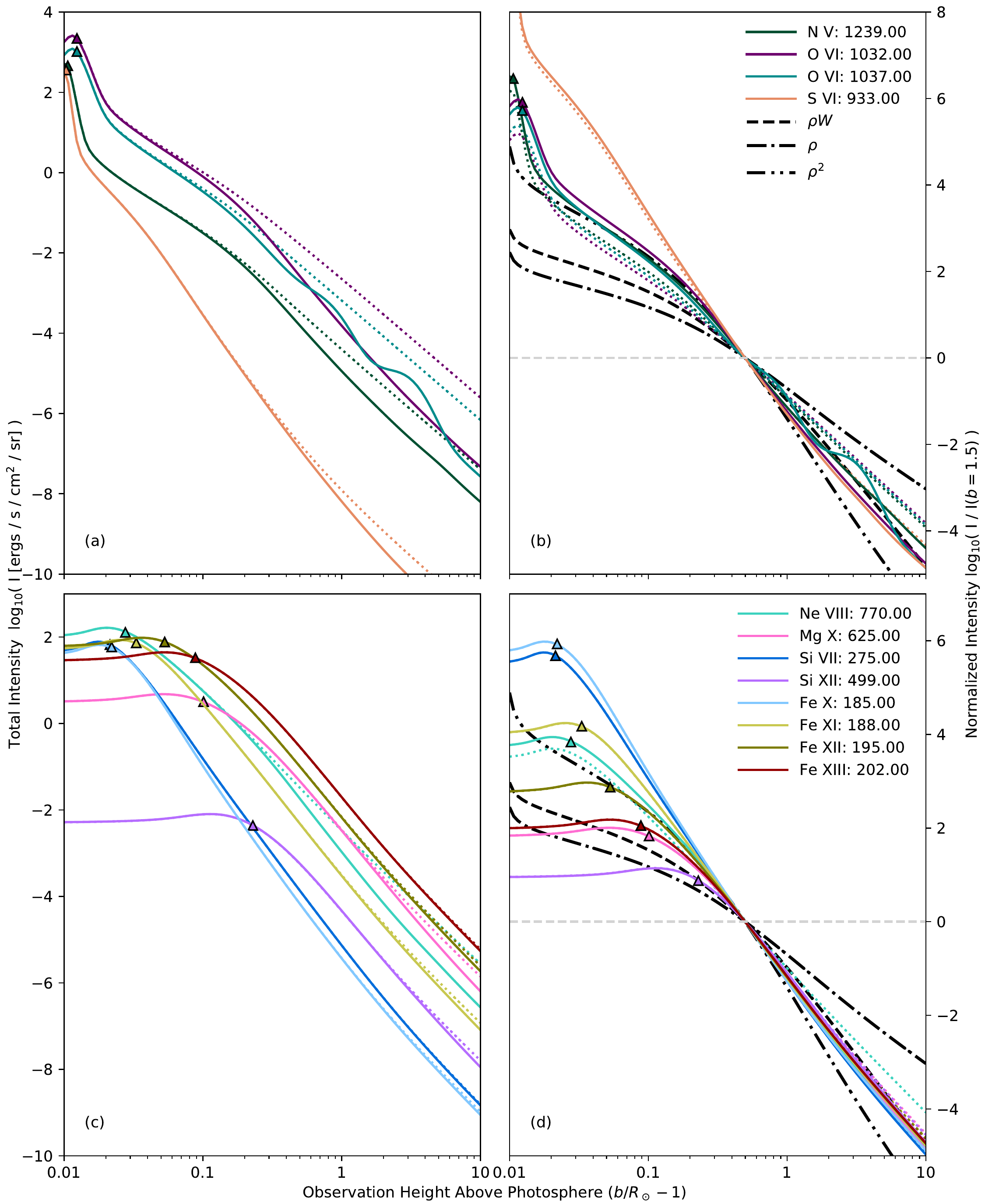}
	\caption{Panels (a) and (c) are in absolute units, and Panels (b) and (d) are normalized. Solid lines show full wind $B=1$, dotted lines show now wind $B=0$. Black lines show the mass density $\rho$ and $\rho^2$. Triangles mark $R_p$ for each ion.}
	\label{fig:cvr-wind2}
\end{figure}

Figure \ref{fig:cvr-wind2} shows the total intensity $I$ for each of the modeled ions. The left column shows absolute units, while the right shows curves normalized to their values at $b=1.5$, where the resonant and collisional intensities are approximately of the same order for many ions. For illustration, we split these ions in to two groups, based on their behavior. Figure \ref{fig:cvr-wind2}(a) and Figure \ref{fig:cvr-wind2}(b) show intensity results for those ions with low peak radii $R_p$ (marked as triangles), which we would expect to behave straightforwardly. The intensities seem to decrease mostly monotonically with height, following $\rho^2$ in the low corona (though \ion{S}{6} 933 drops off much more quickly), and $\rho W$ in the upper corona. It makes sense that they would tend to match these curves, as the lines are more collisionally dominated in the low corona and more resonantly dominated in the upper regions. Comparing the solid lines in Figure \ref{fig:cvr-wind2}(a), which represent a full $B=1$ solar wind case, with the dotted lines showing the flow-free $B=0$ case, one can see that the measurements from the upper corona are dimmed by the presence of the solar wind, but the lower corona is mostly unaffected.

Figure \ref{fig:cvr-wind2}(c) and Figure \ref{fig:cvr-wind2}(d) show intensity results for the remaining ions. These measurements show significant plateaus in intensity, as we would expect from their relatively high $R_p$. The solar wind appears to affect this population to a lesser degree than the cases with lower $R_p$ for the most part.

\begin{figure*}[ht!]
 	\centering
 	\includegraphics[width=\textwidth]{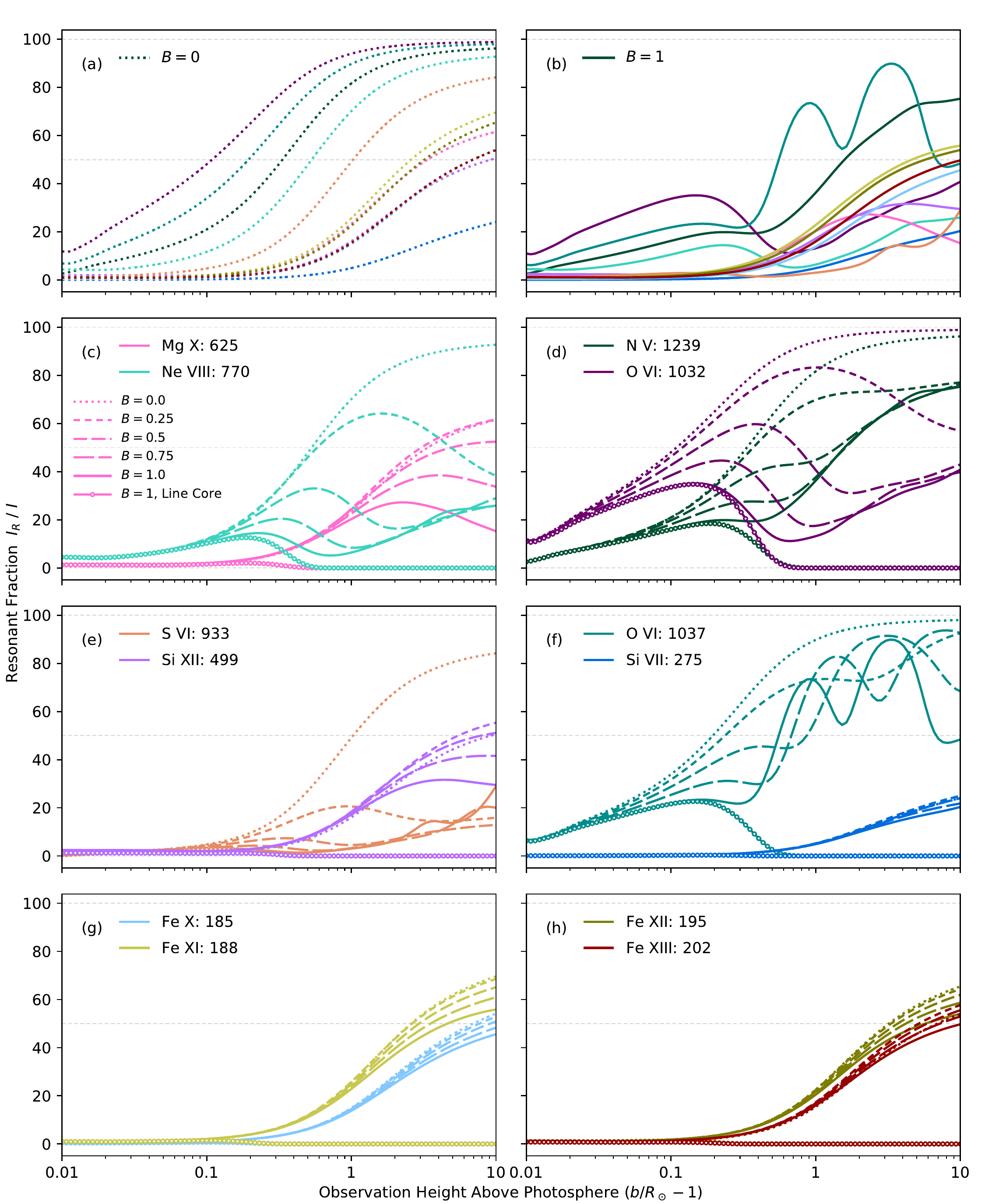}
 	\caption{Contribution of the resonantly scattered photons $I_R$ to the total intensity $I$, with varying solar wind strength. In Panels (a) and (b), all the simulated lines are shown at $B=0$ and $B=1$, respectively. Remaining panels show how each line's behavior is dependent on the wind strength $B$. Line with bubbled style indicates the case with overly-truncated incident light.}
 	\label{fig:cvr-wind}
 \end{figure*}

 Figure \ref{fig:cvr-wind} shows $I_R/I$, the proportion of the total intensity $I$ that is contributed by resonant scattering $I_R$. In general, increasing the strength of the solar wind dims the resonant component of the line due to Doppler dimming. Notable exceptions are the \ion{O}{6} 1037 line, which is Doppler pumped significantly, and \ion{Si}{12} 499, which has slight Doppler pumping. The \ion{O}{6} 1032 line may be a good choice as a wind speed diagnostic, especially above $b=1.5$, as the resonant fraction is strongly inversely proportional to the wind speed $B$.
 
%%%%   Bibliography   %%%%%%%%%%%%%%%%%%%%%%%%%%%%%%%%%%%%%%%%%%%%%%%%%%%%%%%%%%%%%%%%%%%%%%%%%%%%%%%%%%%%%%%%%%%%%%%%%%%%%%

\clearpage
\bibliographystyle{aasjournal}

\end{document}